 \documentclass[twocolumn]{aastex62}
\usepackage{booktabs}
\usepackage{multirow}
\usepackage{tabularx}
\usepackage{graphicx}
\usepackage{amsmath}

\hypersetup{linkcolor=red,citecolor=blue,filecolor=cyan,urlcolor=magenta}

\graphicspath{{./}{figures/}}

\received{May, 2020}
\revised{Month, 2020}
\accepted{Month, 2020}

\submitjournal{ApJ}
\shorttitle{A Universal Fundamental Plane with CALIFA and MaNGA.}
\shortauthors{E. Aquino-Ort\'iz et al.}

\begin{document}

\title{A Universal fundamental plane and the $M_{dyn}-M_{\star}$ relation for galaxies with CALIFA and MaNGA.}

%% The \author command is the same as before except it now takes an optional
%% arguement which is the 16 digit ORCID. The syntax is:
%% \author[xxxx-xxxx-xxxx-xxxx]{Author Name}
%%
%%
%%

\correspondingauthor{E. Aquino-Ort\'iz}
\email{eaquino@astro.unam.mx}

\author[0000-0003-1083-9208]{E. Aquino-Ort\'iz\includegraphics[scale=0.08]{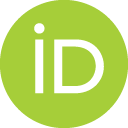}}
\affil{Instituto de Astronom\'ia, Universidad Nacional Aut\'onoma de M\'exico A.P. 70-264, 04510. CDMX, M\'exico}

\author[0000-0001-6444-9307]{S.F. S\'anchez\includegraphics[scale=0.08]{figures/Orcid.png}}
\affil{Instituto de Astronom\'ia, Universidad Nacional Aut\'onoma de M\'exico A.P. 70-264, 04510. CDMX, M\'exico}

\author[0000-0002-0523-5509]{O. Valenzuela\includegraphics[scale=0.08]{figures/Orcid.png}}
\affil{Instituto de Astronom\'ia, Universidad Nacional Aut\'onoma de M\'exico A.P. 70-264, 04510. CDMX, M\'exico}

\author{H. Hern\'andez-Toledo}
\affil{Instituto de Astronom\'ia, Universidad Nacional Aut\'onoma de M\'exico A.P. 70-264, 04510. CDMX, M\'exico}

\author{Yunpeng Jin}
\affil{National Astronomical Observatories, Chinese Academy of Sciences, 20A Datun Road, Chaoyang District, Beijing 100101, China}

\author[0000-0002-8005-0870]{Ling Zhu\includegraphics[scale=0.08]{figures/Orcid.png}}
\affil{Shanghai Astronomical Observatory, Chinese Academy of Sciences, 80 Nandan Road, Shanghai 200030, China.}

\author[0000-0003-4546-7731]{Glenn van de Ven\includegraphics[scale=0.08]{figures/Orcid.png}}
\affil{Department of Astrophysics, University of Vienna, TAijrkenschanzstrasse 17, 1180 Vienna, Austria}

\author[0000-0003-2405-7258]{J.K. Barrera-Ballesteros\includegraphics[scale=0.08]{figures/Orcid.png}}
\affil{Instituto de Astronom\'ia, Universidad Nacional Aut\'onoma de M\'exico A.P. 70-264, 04510. CDMX, M\'exico}

\author[0000-0002-3461-2342]{V. Avila-Reese\includegraphics[scale=0.08]{figures/Orcid.png}}
\affil{Instituto de Astronom\'ia, Universidad Nacional Aut\'onoma de M\'exico A.P. 70-264, 04510. CDMX, M\'exico}

\author[0000-0002-0170-5358]{A. Rodr\'iguez-Puebla\includegraphics[scale=0.08]{figures/Orcid.png}}
\affil{Instituto de Astronom\'ia, Universidad Nacional Aut\'onoma de M\'exico A.P. 70-264, 04510. CDMX, M\'exico}

\author[0000-0001-5242-2844]{Patricia B. Tissera\includegraphics[scale=0.08]{figures/Orcid.png}}
\affiliation{Departamento de Ciencias F\'isicas, Universidad Andr\'es Bello, 700 Fern\'andez Concha, Las Condes, Santiago, Chile.}
%\collaboration{(AAS Journals Data Scientists collaboration)}

\begin{abstract}

We use the stellar kinematics for $2458$ galaxies from the Mapping Nearby Galaxies at Apache point observatory (MaNGA) survey to explore dynamical scaling relations between the stellar mass $M_{\star}$, and the total velocity parameter at the effective radius, $R_e$, defined as $S_{K}^{2}=KV_{R_e}^{2}+\sigma_{\star_e}^{2}$, which combines rotation velocity $V_{R_e}$, and velocity dispersion $\sigma_{\star_e}$. We confirm that spheroidal and spiral galaxies follow the same $M_{\star}-S_{0.5}$ scaling relation with lower scatter than the $M_{\star}-V_{R_e}$ and $M_{\star}-\sigma_{\star_e}$ ones. We also explore a more general two-dimensional surface known as Universal Fundamental Plane described by the equation $log(\Upsilon_{e}) = log (S_{0.5}^{2}) - log (I_{e}) - log (R_{e}) + C.$, which in addition to kinematics, $S_{0.5}$, and effective radius, $R_e$, it includes information of the surface brightness, $I_e$, and dynamical mass-to-light ratio, $\Upsilon_e$. We use sophisticated Schwarzschild orbit-based dynamical models for a sub-sample of 300 galaxies from the CALIFA survey to calibrate the so called Universal Fundamental Plane. That calibration allows us to propose both: (i) a parametrization to estimate the difficult-to-measure dynamical mass-to-light ratio at the effective radius of galaxies, once the internal kinematics, surface brightness and effective radius are known; and (ii) a new dynamical mass proxy consistent with dynamical models within $0.09\ dex$. We show that this dynamical mass estimator is more robust that the one previously proposed using only kinematics. We are able to reproduce the relation between the dynamical mass and the stellar mass in the inner regions of galaxies with lower scatter. We use the estimated dynamical mass-to-light ratio from our analysis, $\Upsilon_{e}^{fit}$, to explore the Universal Fundamental Plane with the MaNGA data set. We find that all classes of galaxies, from spheroids to disks, follow this Universal Fundamental Plane with a scatter significantly smaller $(0.05\ dex)$ than the one reported for the $M_{\star}-S_{0.5}$ relation $(0.1\ dex)$, the Fundamental Plane $(\sim 0.09\ dex)$ and comparable with Tully-Fisher studies $(\sim 0.05\ dex)$, but for a wider range of galaxy types. 

\end{abstract}


\section{Introduction}
\label{sec:Intro}

Observational and theoretical studies of galaxies have revealed the existence of tight correlations between their global stellar and dynamical properties. These correlations reflect the physical connection between photometric properties of galaxies (given by their stellar populations) and their internal kinematics (given by the main dynamical property, the gravitational potential). They illustrate how the gravitational potential (or equivalently the dynamical mass) plays an important role in our understanding of galaxy formation and evolution  \citep[e.g.][]{cole1994,Mo&Mao&White1998,Firmani2000,Courteau2007,Trayford2019}. For example, the empirical Tully-Fisher relation \citep[][hereafter TF]{Tully-Fisher1977} describes a tight correlation between the rotation velocity of spiral galaxies and the stellar mass (or luminosity), with a scatter of $\sim 0.05\ dex$ in velocity \citep[e.g.][]{Verheijen2001,Avila-Reese2008,Reyes2011,Bekeraite2016,Ponomareva2017,Aquino-Ortiz2018} with a break down for velocities smaller than $\sim100\ kms^{-1}$ \citep[e.g.][]{McGaugh2000}. For decades, astronomers have been looking for a third parameter on the TF relation to reduce the scatter \citep[e.g.][]{Zwaan1995,Courteau&Rix1999,Pizagno2007,Avila-Reese2008,Hall2012,Zaritsky2014,Tonini2014}. It appears that no relation tighter than the TF relation can be constructed by including additional information \citep[e.g.][]{Mayer2008}. The analogue to the TF relation for ellipticals is the Faber-Jackson relation \citep[][hereafter FJ]{Faber-Jackson1976}, a correlation between the central velocity dispersion and their total stellar mass (or luminosity) with a scatter of $\sim0.07\ dex$ in velocity dispersion \citep[e.g.][]{Gallazzi2006}. Unlike the TF relation, there is a third parameter that generates a tighter correlation than the FJ relation. That is a correlation among the central velocity dispersion, $\sigma_{\star_e}$, the effective radius, $R_{e}$, and the average surface brightness at the effective radius, $I_{e}$. This relation is called the Fundamental Plane (hereafter FP) \citep{Djorgovski&Davis1987,Dressler1987}. Fitting a plane to the data ($I_{e}$, $\sigma_{\star_e}$, $R_{e}$) yields three coefficients assuming the functional form $log(R_{e}) = a\ log(\sigma_{\star_e}) + b\ log(I_{e}) + c $. From the virial theorem and assuming that the elliptical galaxies have: (i) constant mass-to-light ratios, $M/L$, (ii) structure spherically symmetric, (iii) dynamically homologous density and orbital profiles, and (iv) similar dark matter fractions, then a FP with a = 2 and b = −1 is expected. However, a deviation from the virial prediction, called the tilt of the FP, derives $a = 1.063 \pm 0.041$ and $b = −0.765 \pm 0.023$ with a scatter of $\sim0.09\ dex$ in $log(R_{e})$ \citep{Cappellari2013}.
The origin of the tilt have been attributed to deviations of the above assumptions. For example: (i) variations in the $M/L$ values increasing systematically with luminosity \citep[e.g.][]{Faber1987,Garcia-Benito2019}, (ii) variations in the kinematic and density profiles \citep[e.g.][]{Prugniel&Simien1994,Graham&Colless1997,Busarello1997,Bertin2002,Trujillo2004}, (iii) variations in the stellar populations \citep[e.g,][]{Sebastian2019Review} or initial mass function \citep[e.g.][]{Prugniel&Simien1996,Forbes1998,Dutton2013,Martin-Navarro2015} and (iv) variations in dark matter fraction \citep[e.g.][]{Renzini1993,Ciotti1996,Borriello2003,Padmanabhan2004}. An extension of the FP called Fundamental Manifold (hereafter FM) was introduced by \citet[][]{Zaritsky2006} for spheroidal dominated stellar systems. They included the efficiency with which baryons are packed with respect to dark matter measured by the dynamical mass-to-light ratio within $R_e$, $(\Upsilon_{e} = M_{dyn_e}/L_e)$, to define the FM as $log (R_{e}) = 2\ log (\sigma_{\star_e}) - log (I_{e}) - log(\Upsilon_{e}) + C $. They found a scatter to the FM of $\sim0.1\ dex$ in $log(R_{e})$, similar to that of the FP.%, both for 
%elliptical galaxies.

As we have discussed so far, these scaling relations work just over a limited range of galaxy types, suggesting that the current scaling laws are incomplete. It is not yet evident that any of these relations are as fundamental for all galaxy types as the main sequence on the Hertzsprung-Russell diagram is for stars. Previous studies have tried to unify kinematic scaling relations for spiral and elliptical galaxies. For example, \citet[][]{Falcon-Barroso2011} explored the FP for a representative sample of 72 galaxies including Sa galaxies from the SAURON survey \citep{Bacon2001}. They find a FP with the lowest scatter defined by the Slow Rotators galaxies ($\sim 0.062\ dex$), whereas the Fast Rotators display a slightly larger
scatter ($\sim 0.081\ dex$). The scatter for the Sa galaxies appears to be the largest ($\sim 0.165\ dex$). \cite{Weiner2006} introduced a kinematic parameter which combines the rotation velocity at $R_e$, $V_{R_e}$, and central velocity dispersion, $\sigma_{\star_e}$. The parameter was defined as:

\begin{equation}
    S_{K}^{2} = KV_{R_e}^{2} + \sigma_{\star_e}^{2},
\end{equation}
where $K$ generally is assumed to be constant \citep[e.g.][]{Kassin2007ApJ,DeRossi2012}. However, it could be different for each galaxy and/or be a complicated function of different galaxy properties like the formation and evolution history, dynamical state, environment, etc.

\citet[][]{Cortese2014} \& \citet[][]{Aquino-Ortiz2018} for the Sydney-AAO (Australian Astronomical Observatory) Multi-object IFS \citep[SAMI,][]{Croom2012MNRAS} and the Calar Alto Legacy Integral Field Area \citep[CALIFA,][]{SanchezCALIFA2012A&A} surveys respectively, showed that all galaxies, regardless of the morphological type, lie on the same $M_{\star}-S_{K}$ scaling relation with a scatter of $\sim 0.1\ dex$ in $log(S_{K})$ (smaller or equal to that for the $M_{\star}-V_{R_e}$ and $M_{\star}-\sigma_{\star_e}$ relations). Different studies have found that the fitted $M_{\star}-S_{K}$ relation reaches its minimum scatter for K = 0.5 \citep[e.g.][]{Aquino-Ortiz2018,Gilhuly2019,Barat2019MNRAS}. This total velocity parameter, $S_{0.5}$, has been used in a General Fundamental Manifold for galaxies (hereafter GFM), for Early Type Galaxies (hereafter ETGs) and Late Type Galaxies (hereafter LTGs), firstly by \citet[][]{Zaritsky2008} and refined in \citet[][]{Zaritsky2012}. They provide a fitting function for the dynamical mass-to-light ratio within $R_{e}$, $log\ (\Upsilon_{e})$, that depends only on $S_{K}$ and $I_{e}$. They show that all classes of systems, from spheroids to disks, fall on the GFM with a scatter of $\sim 0.1\ dex$ in $log(R_e)$, comparable to that observed in the FP studies \citep[e.g.][]{Cappellari2013} and FM by \citet[][]{Zaritsky2006} (in which the range of galaxy types is limited to ETGs). They also found that the GFM for sub-samples with independently derived dynamical mass-to-light ratios suggest a intrinsic scatter as low as $\sim 0.05\ dex$. More recently, \citet[][]{Li2018MNRAS} explored the Mass Plane relation for about 2000 galaxies from the MaNGA survey. They found that LTGs and ETGs follow this tight mass plane with a observed scatter of $\sim 0.06\ dex$ and $\sim 0.04\ dex$, respectively. A scatter slightly larger for LTGs.

Understanding the origin and evolution of galaxies remains the principal goal behind refining and understanding the scaling relations discussed so far. However, they also provide numerous practical/useful benefits. For example, they have been used as distance estimators \citep[e.g.][]{Giovanelli1997,Zaritsky2012ApJDistances}, or as proxies of the galaxy dynamical mass \citep[e.g.][]{Courteau2014,Aquino-Ortiz2018}. With this in mind, in this work we are focused in two main goals: (a) explore the remarkably tight $M_{\star}-S_{K}$ scaling relation using the Mapping Nearby Galaxies at APO \citep[MaNGA,][]{Bundy2015} data set; and (b) calibrate the so called Universal Fundamental Plane (hereafter UFP) proposed by \citet[][]{Zaritsky2008}, valid for early and late type galaxies. For the calibration we use a representative sub-sample of 300 galaxies for from the CALIFA survey with independently derived dynamical mass-to-light ratios, $\Upsilon_e$. Finally, we want to provide reliable estimations of the dynamical masses based on the Universal Fundamental Plane and explore the $M_{dyn_e}-M_{\star_e}$ relation.

In Section \ref{sec:sample} we briefly describe the CALIFA and MaNGA samples. Details of the analysis performed over the data are presented in Section \ref{sec:Analysis}. In Section \ref{PIPE3D_Analisys} we describe the stellar population synthesis applied to the data. In Section \ref{subsec:sample_selec} we describe the sample selection. In Section \ref{subsec:Int_kin} we present the integrated kinematic analysis within $R_{e}$. In Section \ref{SRK} we perform a detailed modelling of the 2D spatially resolved velocity maps for a sub-sample of good quality data sets to obtain a more precise derivation of the maximum rotational velocity $(V_{max})$. In Section \ref{sec:Results}, we present the main results, while in Section \ref{sec:mass_estimator} we present our dynamical mass estimator for galaxies and explore the $M_{dyn_e}-M_{\star_e}$ relation. In Section \ref{Discussion} we discuss the physical implications of our main results. Finally we summarize the main conclusions in Section \ref{Summary}.
Throughout this article we adopt a cosmology with $H_{0}=70kms^{-1}Mpc^{-1}$, $\Omega_{M} = 0.3$ and $\Omega_{\Lambda} = 0.7$ for the Hubble constant, the matter density and the cosmological constant, respectively.

\section{Data Sample} \label{sec:sample}

This study is based on data provided by the CALIFA \citep{SanchezCALIFA2012A&A} and MaNGA \citep{Bundy2015} surveys, particularly the 4676 galaxies from the MaNGA Product Launch-7 (MPL-7) publicly available since June 2018. In this section we briefly describe each of these surveys.

\begin{figure*}[ht]
%    \begin{subfigure}{.3\textwidth}
    \centering
%  % include first image
    \includegraphics[width=0.49\linewidth]{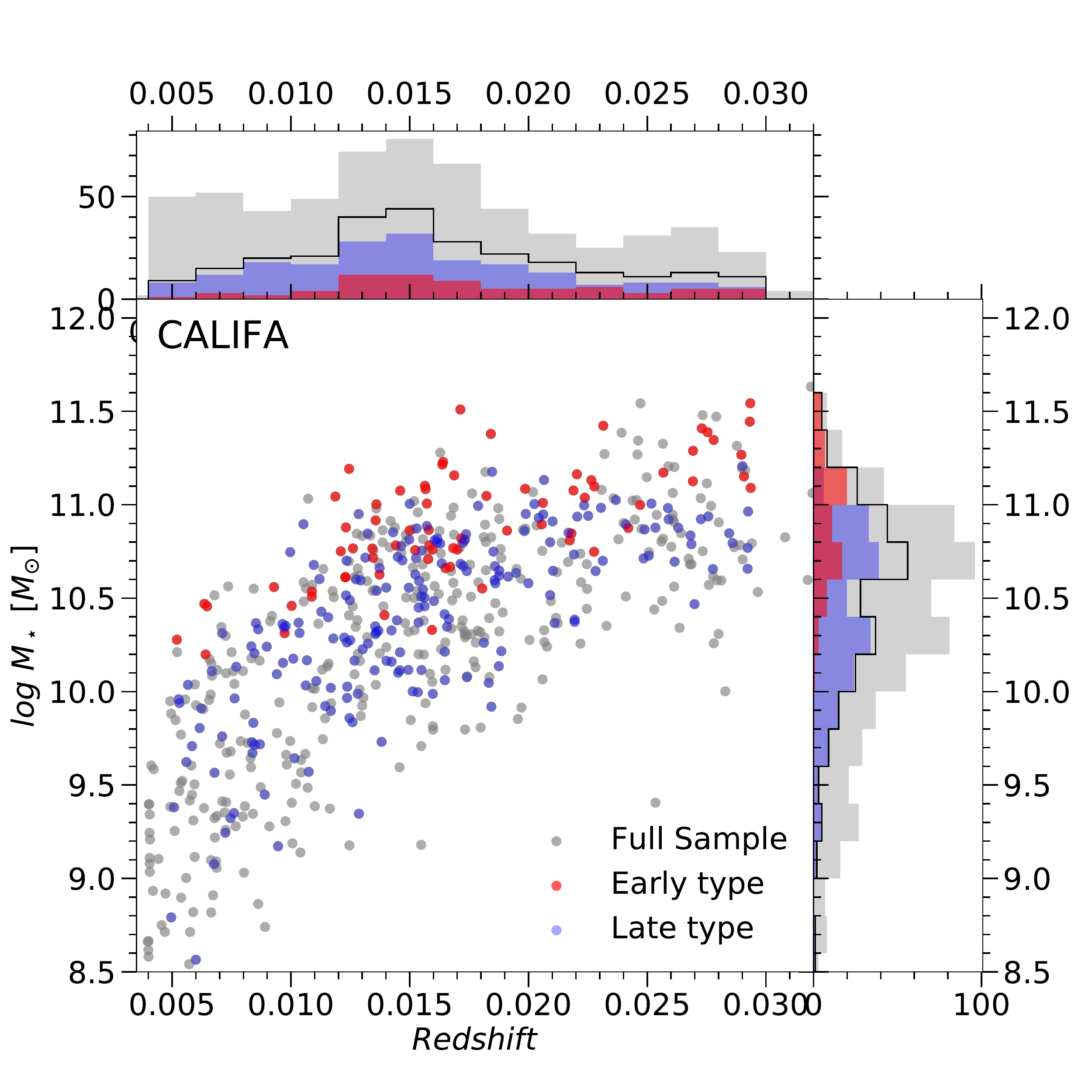}
    \includegraphics[width=0.49\linewidth]{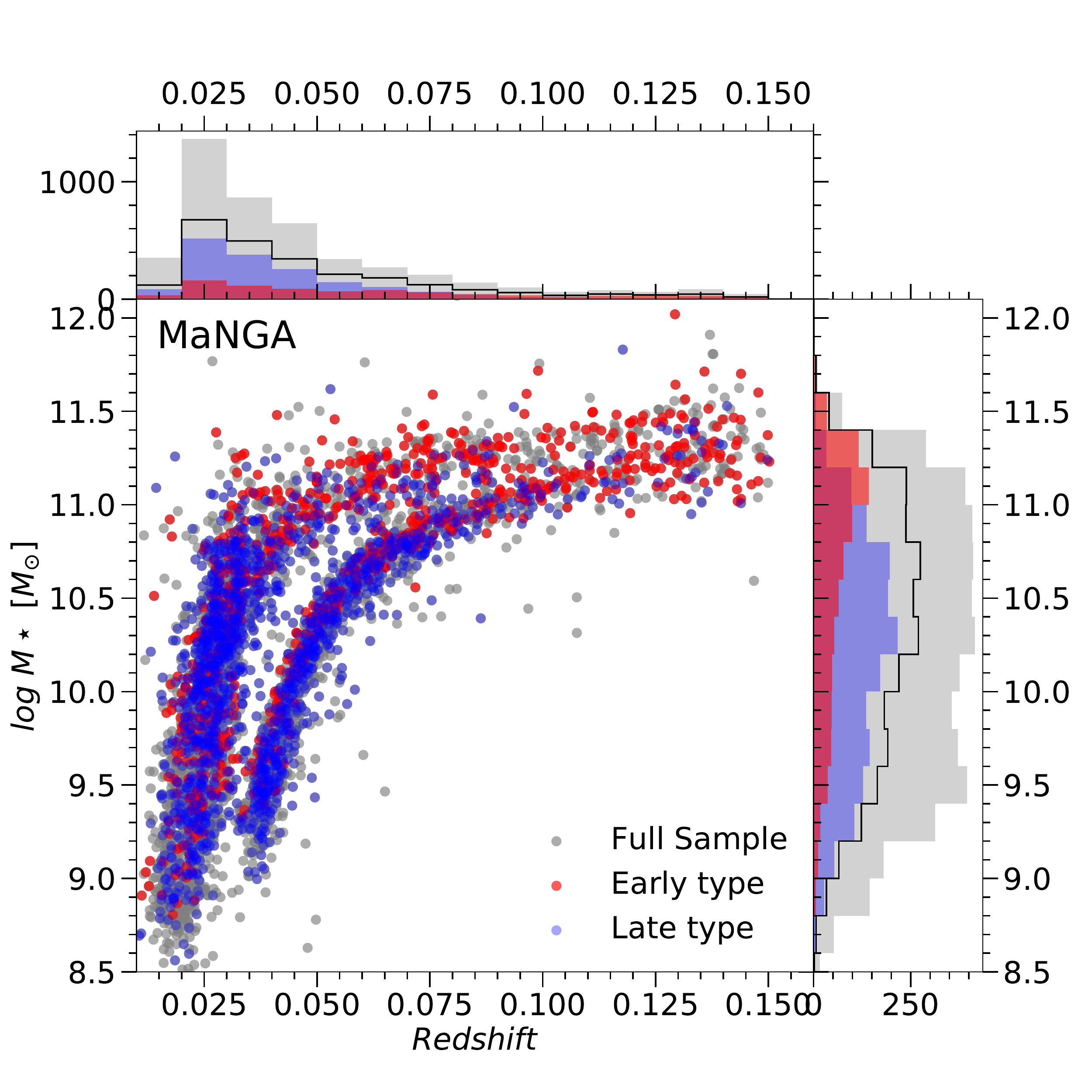}
\caption{Sample selection boundaries. \textit{Left panel:} CALIFA Sample. \textit{Right panel:} Primary and Secondary MaNGA Samples. The gray symbols and histograms represents the full CALIFA sample of 667 galaxies (left panel) and full MaNGA MPL-7 data set (right panel). In both panels blue and red symbols and histograms represents LTGs and ETGs used in this study. The black histogram is the sample of early+late types.}
\label{fig:MaNGA_sample}
\end{figure*}

\subsection{The CALIFA survey}
\label{sec:CALIFA_DATA}

One of the aims of this paper is to calibrate the Universal Fundamental Plane. To do so we use the publicly available data provided by the CALIFA survey \citep{SanchezCALIFA2012A&A}. CALIFA observed a statistically representative sample of 667 galaxies of all morphological types and environments \citep{Sanches3DR2016A&A}, recently increased by a set of extended complementary observations comprising $\sim 900$ objects \citep[e.g.][]{Lacerda2020MNRAS,Espinosa-Ponce2020}. The galaxies were selected to have a major axis diameter $45" < D_{25} < 80"$, where $D_{25}$ is the isophote major axis at $25\ mag/arcsec^{2}$ in the SDSS r-band. The sample comprises galaxies in the local universe ($0.005 < z < 0.03$) in a stellar mass range of $10^{8.5}<M_{\star}<10^{11.5}M_{\odot}$ \citep[See left panel of Figure \ref{fig:MaNGA_sample}. Further details on the galaxy sample selection see][]{Walcher2014A&A}. The galaxies were observed with the Potsdam Multi Aperture Spectrograph \citep[PMAS,][]{Roth2005PASP} in the PPaK configuration. The PPak system consists of a fiber bundle with 331 object fibers, 36 sky fibers, and 15 calibration fibers ($2.7"$ in diameter each one). That configuration covers an hexagonal field of view (FoV) of $74"$ x $64"$, sufficient to map the full optical extension of most of the galaxies up to 2-3 $R_e$ \citep{Kelz2006PASP}. The median spatial resolution is FWHM$\sim2.5"$ that corresponds on an average physical resolution of $0.8$ kpc \citep{GarciaBenito2015A&A}. Observations were carried out in two configurations: (i) the V500 setup, a low resolution mode ($\lambda/\Delta\lambda\sim 850$ at $\sim5000$\AA, corresponding to $\sigma_{inst}\sim150\ km/s$), covering the spectral range between 3750 and 7500\AA, and (ii) the V1200 setup, an intermediate resolution mode ($\lambda/\Delta\lambda\sim1650$ at $\sim4500$\AA, corresponding to $\sigma_{inst}\sim70\ km/s$), covering the wavelength range between 3700 and 4800 \AA.
The data set was reduced with the version 2.2 of the CALIFA pipeline, whose improvements with respect to the previous ones \citep{SanchezCALIFA2012A&A,Husemann2013A&A,GarciaBenito2015A&A} are reported in \citet{Sanches3DR2016A&A}. The final data-product after the reduction is a data-cube with the spatial information along the x- and y-axis, and the spectral one in the z one.

For the calibration to the UFP in this study, we use the properties for the sub-sample of 300 CALIFA galaxies covering all galaxy types from the V1200 setup presented in \citet{Falcon-Barroso2017}. The galaxy properties used here for each galaxy in that sub-sample are: (i) the publicly available\footnote{Publicly available stellar line-of-sight kinematic maps in \url{http://califa.caha.es/?q=content/science-dataproducts}} stellar velocity and velocity dispersion maps computed by \citet{Falcon-Barroso2017} using the pPXF code of \citet{CappellariPPXF2004PASP}. We use these kinematic maps to estimate the total velocity parameter, $S_{0.5}$; (ii) apparent magnitudes; and (iii) effective radius estimated using a growth-curve analysis applied to the r-band SDSS images by \citet{Walcher2014A&A}. We use these properties to estimate the luminosity at $R_{e}$, $L_{e}$, hence the surface brigthness, $I_{e}$. Finally, (iv) the dynamical mass-to-light ratios at $R_{e}$, $\Upsilon_{e}^{Sch}=M_{dyn_e}/L_e$, derived by \citet{Zhu2018} through a full Schwarzschild orbit-based dynamical technique \citep{Schwarzschild1979}.

\subsection{The MaNGA survey}
\label{sec:MaNGA_DATA}
The MaNGA survey \citep{Bundy2015} began in July 2014 as part of the Sloan Digital Sky Survey-IV collaboration \citep[SDSS-IV,][]{Blanton2017}. The aim of MaNGA is observe a sample of $10.000$ galaxies with the integral field spectroscopy technique over a broad wavelength range $(3600-10300\ $\AA). Observations are performed using the SDSS 2.5 meters telescope at Apache Point Observatory \citep[APO,][]{Gunn2006} and the SDSS-III Baryonic Oscillation Spectroscopic Survey spectrograph \citep[BOSS,][]{Smee2013}. The resolving power is $R = \lambda/FWHM \sim 2000$, i.e., $\sigma_{inst}  \sim 70\ km/s$. MaNGA deploys a set of 17 Integral Field Units (IFU) grouped into hexagonal bundles of different sizes ranging from 19 to 127 optical fibers of 2" in diameter each one \citep{Drory2015}. The observations are dithered adopting a three-point triangular pattern on the sky to achieve a complete spatial coverage of the sources \citep{Law2015}. The sample comprising galaxies of any morphological type and environments was chosen in a redshift range of $0.01 < z < 0.15$ with approximately flat stellar mass distribution with  $M_{\star} \geq 10^{8.5}\ M_\odot$ \citep[See right panel of Figure \ref{fig:MaNGA_sample}. For further details about the sample design see][]{Wake2017}. With a median spatial resolution of 1.8 kpc, the main MaNGA sample consists of three components: (i) the Primary sample covered out to $1.5 R_e$ representing the $47\%$ of the main sample. It is selected so that $80\%$ of the galaxies in this Primary sample can be observed with the 127 fiber-bundle, (ii) the Secondary sample representing the $37\%$ of the main sample covered out to $2.5 R_e$ is designed to observe the $80\%$ of galaxies with the bundle of 127 fibers, and (iii) the Color-Enhanced which represents the $16\%$ of the main sample includes low luminosity red galaxies, high luminosity blue galaxies and green valley galaxies to fill in poorly sampled regions of the $NUV-i\ vs\ M_i$ color-magnitude diagram. About $5\%$ of all MaNGA galaxies are selected from different ancillary programs addressing several scientific goals.

The Data Reduction Pipeline \citep[DRP;][]{Law2016} reduces the single fibers in each exposure into sky subtracted, wavelength and flux calibrated individual spectra. The final data-product of the reduction is a three-dimensional data-cube that combines individual dithered observations comprising the spatial information in the $x$- and $y$-axis, and the spectral one in the $z$-axis.

Along this study, for the MaNGA data set we use the following galaxy properties: (i) the line-of-sight kinematic maps and mass distribution derived by the PIPE3D data-products described in the next section, (ii) the effective radius, $R_{e}$, and total stellar mass, $M_{\star}$, extracted from the NSA catalog \citep[][ \url{http://www.nsatlas.org/}]{Blanton2011}, (iii) the r-band apparent magnitudes from \citet{Fischer2019MNRAS} to compute the luminosity $L_{e}$, hence the surface brightness $I_{e}$.

We also benefit from a detailed visual morphological classification to 4676 MaNGA galaxies in MPL-7,  based on a new reprocessing of the SDSS images in combination with additional image processing to the Dark Energy Legacy Survey images \citep[DESI,][]{DESI2016}. This new classification will be presented elsewhere (Vazquez-Mata et al. in prep.)
\section{Analysis.} \label{sec:Analysis}

We describe in this Section the analysis performed to estimate the stellar mean velocity and velocity dispersion maps as well as the stellar mass distribution for the MaNGA galaxies.

\subsection{Spectroscopic analysis.}
\label{PIPE3D_Analisys}

In the current study we use the MaNGA data-products which are part of the Pipe3D Valued Added Catalog (VAC) included in the DR15\footnote{The Pipe3D VAC included in the DR15 is accessible at: \url{https://www.sdss.org/dr15/manga/manga-data/manga-pipe3d-value-added-catalog/}}. The Pipe3D pipeline \citep{Sanchez2016a,Sanchez2016b,Sanchez2018AGN} was developed to perform the spatially resolved stellar population analysis of the data cubes. Pipe3D applies a spatial binning to the data cubes with the goal of reaching a homogeneous signal-to-noise (S/N) of 50 across the Field of View (FoV). After that, it models the stellar continuum for each co-added spectra within each spatial bin adopting a multi-Single Stellar Population (SSP) template library, taking into account stellar velocity, velocity dispersion, and dust attenuation of the stellar population. The GSD156 template library described in detail by \citet{Cid_Fernandes2013} was adopted with a Salpeter Initial Mass Function \citep{Salpeter1955}\footnote{All stellar masses, $M_{\star}$, along this study are converted from \citet{Salpeter1955} to \citet{Chabrier2003} Initial Mass Function.}. This library comprises 156 templates covering 39 stellar ages (from 1 Myr to 13 Gyr) and four metallicities ($Z/Z_{\odot}=0.2$, 0.4, 1.0 and 1.5). Following to \citet{Cid_Fernandes2013} and \citet{Sanchez2016a} the stellar population model for each spaxel was estimated by rescaling the best-fitted model within each spatial bin to the continuum flux intensity in the corresponding spaxel. The stellar-population model spectra are then subtracted from the original data-cube to create a gas-pure cube. This cube is then used to estimate the main properties of the nebular emission lines.
Additionally, Pipe3D recovers the spatial distribution of the stellar mass densities and the integrated stellar mass at different apertures by taking in account: (i) the decomposition in SSPs, (ii) the stellar mass-to-light ratio for each of them, and (iii) the integrated light at each spaxel within the FoV.

\begin{figure*}[ht]
%    \begin{subfigure}{.3\textwidth}
    \centering
%  % include first image
    \includegraphics[width=0.31\linewidth]{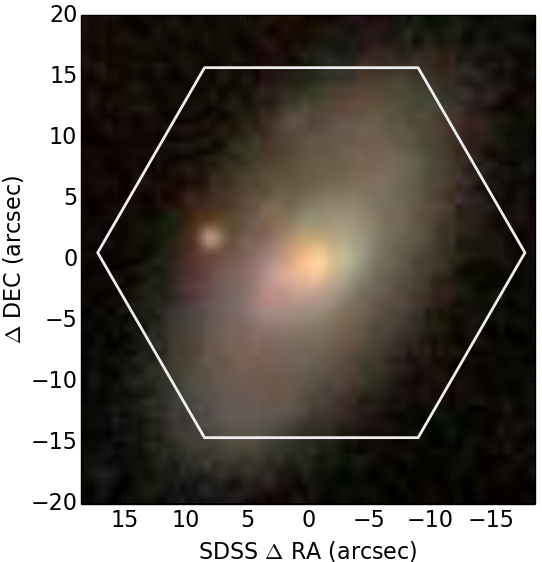}  
%    \caption{MaNGA Field of View.}
%    \label{fig:sub-first}
%    \end{subfigure}
%    \begin{subfigure}{.3\textwidth}
%    \centering
  % include second image
    \includegraphics[width=0.31\linewidth]{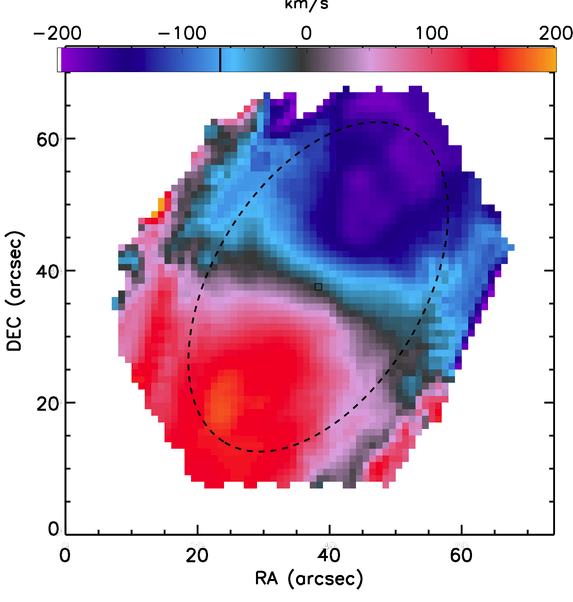}
%    \caption{Stellar velocity map.}
%    \label{fig:sub-second}
%    \end{subfigure}
%    \begin{subfigure}{.3\textwidth}
%    \centering
  % include second image
    \includegraphics[width=0.35\linewidth, height=6cm]{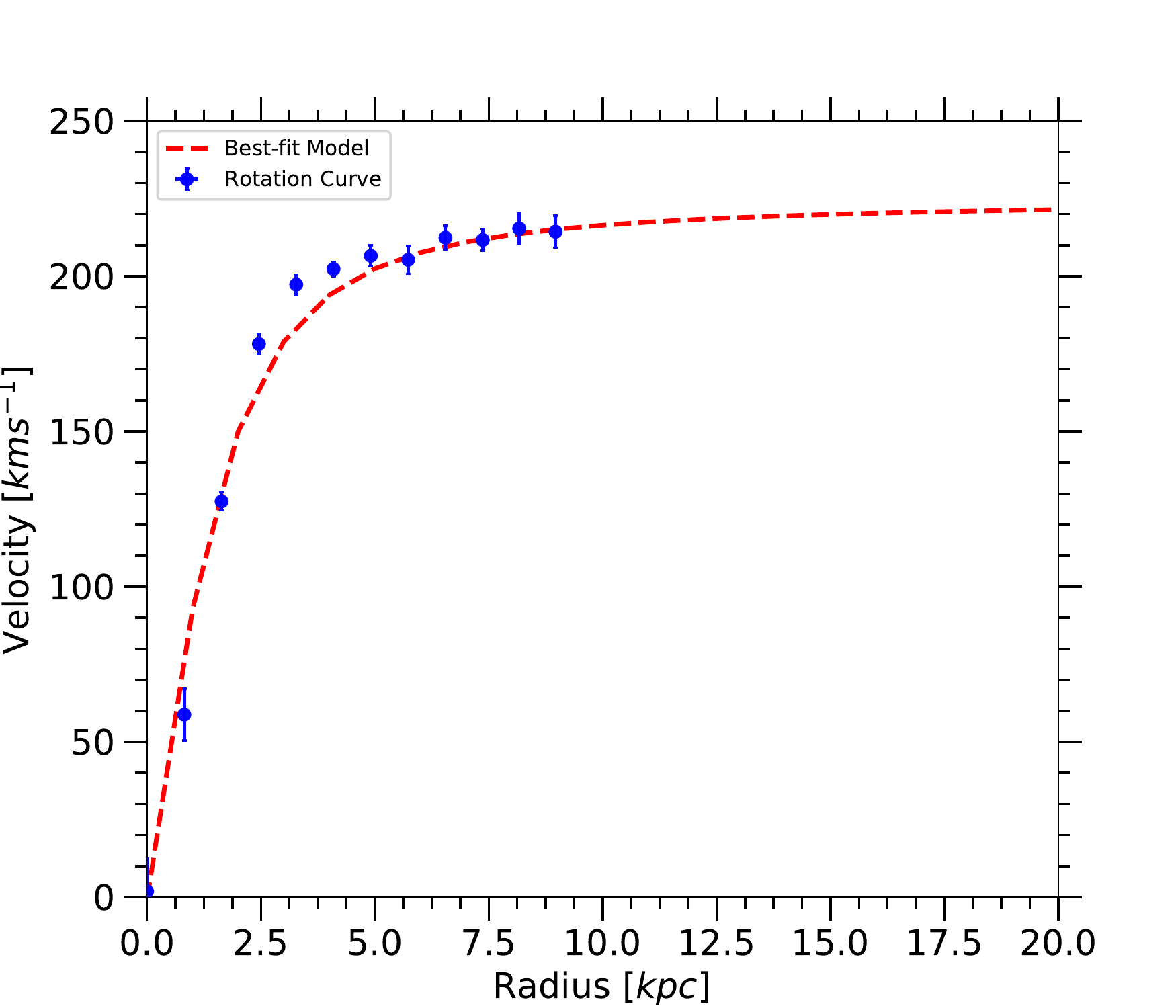}
%    \caption{Rotation Curve.}
%    \label{fig:sub-second}
%    \end{subfigure}
\caption{Example of spatially resolved kinematics from the MaNGA stellar kinematic maps. \textit{Left panel:} the rgb SDSS image and the FoV covered by MaNGA for the manga-8440-12704 galaxy. \textit{Middle panel:} the line-of-sight stellar velocity map. The dashed black ellipse represent the more external ring explored in the analysis. Only the rotation component was modeled ignoring no-circular motions. \textit{Right panel:} the rotation curve derived from the stellar velocity map (blue symbols) with the analysis described in Section \ref{SRK}. The dashed red line represent the best-fit parametrization of Eq. (\ref{eq:RotCurvParam}) to the blue data points.}
\label{fig:VelmapandCurve}
\end{figure*}

\subsection{Sample selection.} \label{subsec:sample_selec}

For the original sample of 4676 galaxies from the MPL-7 data set we perform a selection of the optimal stellar kinematic data for the current analysis following the procedure described in previous studies \citep[e.g.,][]{Cortese2014,Aquino-Ortiz2018}. First, for each stellar velocity and velocity dispersion map we discarded spaxels with errors in velocity larger than 25 km/s. This conventional cut corresponds to $\sim 1/3$ of the average spectral resolution ($\sigma_{inst} \approx 70$ km/s) for the MaNGA data. Second, we select only those galaxies for which at least 60\% of the spaxels within an ellipse of semi-major axis equal to $R_e$ fulfill this quality criterion (using the position angle and inclination of each galaxy). In addition, to minimize the effects of dust, edge-on galaxies with inclinations larger than $75^{\circ}$ were excluded. To derive a reliable rotation measurement, nearly face-on galaxies with inclinations lower than $25^{\circ}$ were removed. Finally, galaxies under merging and clear traces of interactions are discarded based on either morphological distortions, or the presence of galaxy neighbors with a comparable size \citep[e.g.][]{Barrera-Ballesteros2015A&A,Barrera-Ballesteros2015A&Ab}. Following this procedure our final sample for the MPL-7 comprises 2458 galaxies, of which 1653 corresponds to LTGs and 805 to ETGs.
The effects of not making a detailed selection of the galaxies and spaxels within galaxies following this criteria are explored in Appendix \ref{apx:MPL9-dataset}. In particular it is shown the changes in the slopes and zero-points.

\subsection{Integrated kinematics analysis.} \label{subsec:Int_kin}

We derive the stellar mean velocity and velocity dispersion following \citet{Cortese2014} and \citet{Aquino-Ortiz2018}:

\begin{itemize}
    \item \textbf{Velocity dispersion, $\sigma_{\star_e}$}; Integrated velocity dispersions are estimated as the linear average of the velocity dispersion of all "good" spaxels within the ellipse mentioned in the previous section. We use linear instead of luminosity-weighted averages to be consistent with our rotation velocity measurements which are not luminosity-weighted.
    
    \item \textbf{Rotation Velocity, $V_{R_e}$}; We derive the rotation velocities in a similar way to the classical procedure developed to analyze the integrated HI emission profiles in galaxies \citep[e.g.][]{Vogt2004,Papastergis2011}. First, a velocity histogram is derived with the velocities of all the good spaxels within the ellipse defined before. Following \citet[][]{Catinella2005AJ}, the width of the distribution, W, is defined as the difference between the $10^{th}$ and the $90^{th}$ percentile points of the velocity histogram: $W = V_{90} - V_{10}$. Finally, the observed velocity widths derived from the velocity histograms must be corrected for cosmological broadening (to obtain the rest-frame velocities) and de-projected to an edge-on view as follow:

\begin{equation}
V_{R_e} = \frac{W}{2(1+z)sin(i)},
\end{equation}
where $z$ is the redshift and $i$ is the galaxy inclination calculated from the observed ellipticity, $\varepsilon$, as:

\begin{equation}
    cos(i) = \sqrt{\frac{(1 - \varepsilon)^{2} - q_{0}^{2}}{1 - q_{0}^{2}}}
\end{equation}
with $q_{0}$ being the intrinsic axial ratio for galaxies. Following \citet{Rodriguez&Padilla2013} and \citet{Zhuq062018} we adopted $q_{0}=0.2$ for LTGs and $q_{0}=0.6$ for ETGs.
\end{itemize}

\subsection{Spatially resolved kinematics: $V_{max}$} \label{SRK}

As an extension of the integrated kinematics study, we made a detailed analysis for a sub-sample of spiral galaxies to measure $V_{max}$ from rotation curves. Is well know that for a self-gravitating exponential disk the expected maximum velocity of the rotation curve is reached at 2.2 disk scale lengths, $r_{d}$ \citep{Freeman1970}. Since to the gravitational potential of galaxies contribute not only the disk, but also the dark matter halo, which is more extended, then one expect in many cases $V_{max} > V_{2.2}$. Moreover, galaxy discs can show significant deviations from purely exponential profiles \citep[e.g.][]{Bakos2012}, and rotation curves present also a wide range of shapes \citep[e.g.][]{Kalinova2017MNRAS}. To ensure that we reach the $V_{max}$, we select the analyzed sub-set from the Secondary MaNGA sample (FoV $\geq\ 2.5 R_{e}$), and only for the largest MaNGA bundles (i.e., the ones with 127 fibers).

Velocity maps of galaxies frequently exhibit signatures of non-circular streaming motions produced by structural properties, internal physical processes, environment, outflows, inflows, etc. \citep[e.g.][]{Valenzuela2007,Holmes2015}. This effect can be reflected in the shape of the rotation curve and therefore it produces an over/under estimation of $V_{max}$ \citep[e.g.][]{Randriamampandry2015MNRAS}. With this in mind, we choose the sub-sample with a visual inspection to discard those galaxies whose kinematics appeared highly disturbed. Our refined sub-sample comprises 200 galaxies with inclinations in the range of $25^{\circ}< i < 75^{\circ}$.

\begin{figure*}[ht!]
\centering
%\plotone{MaNGA_ScaRel.pdf}
\includegraphics[width=25cm,height=11cm,keepaspectratio]{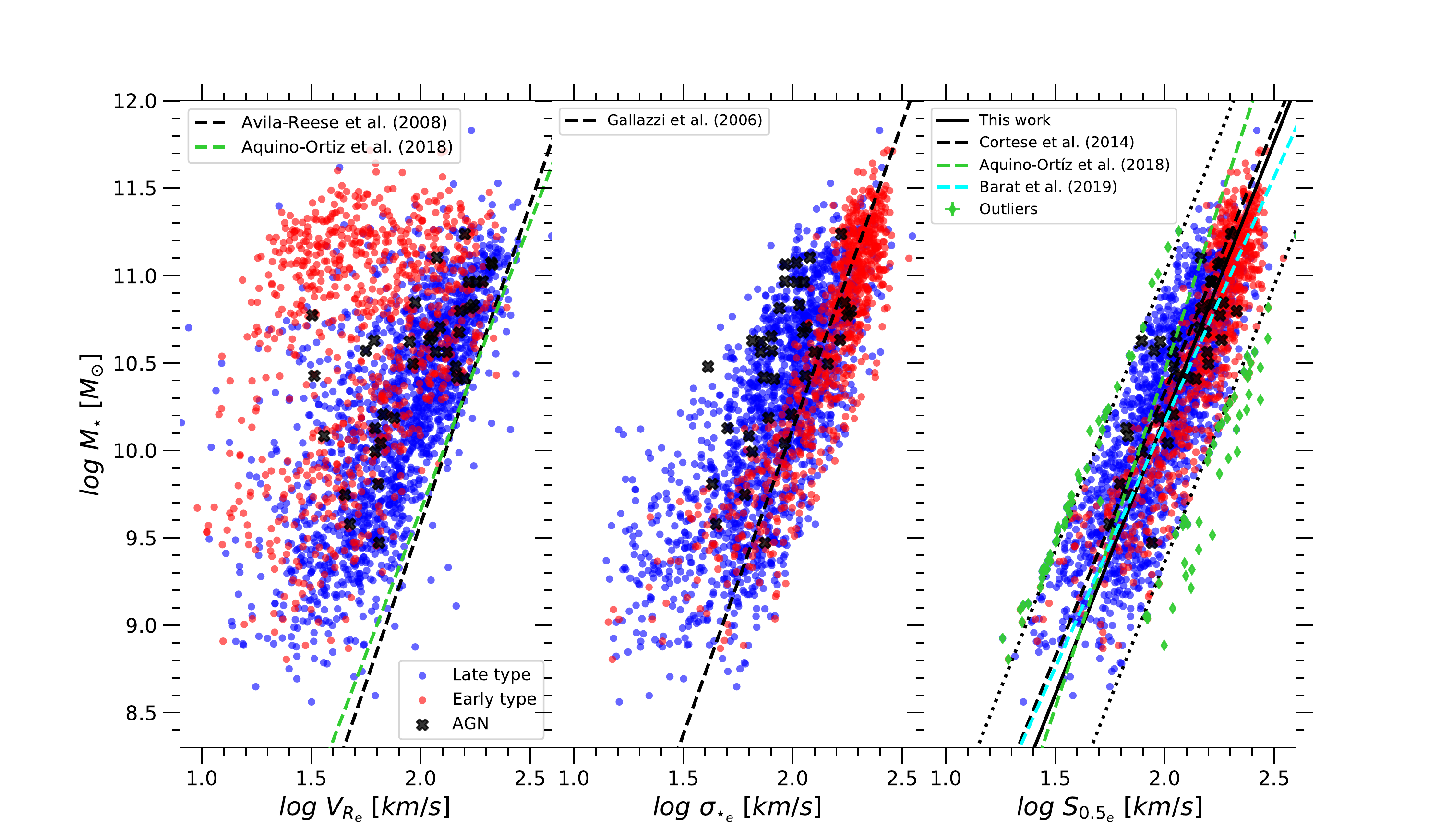}
\caption{Scaling relations with integrated kinematics for the MaNGA data set. Red and blue symbols represent early and late type galaxies, respectively, while black symbols are the AGN sample from \citet{Sanchez2018AGN}. \textit{Left panel:} $M_{\star}-V_{R_e}$ relation; the black and green dashed lines represents the orthogonal best-fits to the classical TF relation from \citet{Avila-Reese2008} and \citet{Aquino-Ortiz2018} using $V_{max}$, respectively. \textit{Middle panel:} $M_{\star}-\sigma_{\star_e}$ relation; the dashed line represents the best-fit for the FJ relation from \citet{Gallazzi2006}. \textit{Right panel:} $M_{\star}-S_{0.5}$ relation; green symbols are the ouliers, black and cyan dashed lines indicate the best-fits from \citet{Cortese2014} \& \citet{Barat2019MNRAS} for the SAMI survey and the green dashed line
the best-fit from \citet{Aquino-Ortiz2018} for the CALIFA one, respectively. The black solid and dotted lines represents our best-fit, and the $2\sigma$ band.} 
\label{fig:Int_kin}
\end{figure*}

Following \citet{Aquino-Ortiz2018}, we use a modified version of the Velfit code (Aquino-Ortiz et al. in prep), which originally was developed by \citet{Spekkens2007} and \citet{Sellwood2010} to make a detailed kinematic analysis (See an example in the Figure \ref{fig:VelmapandCurve}). The purpose of this code is to fit velocity maps of galaxies with a model including a flow pattern in an idealized non-axisymmetric potential. The kinematic model yields: (i) the rotation curve, (ii) the amplitudes of radial and tangential non-circular motions when they are present, and (iii) an estimation of the geometrical parameters without assuming small deviations from circular motions. Once derived the rotation curve, we measure $V_{max}$ by parametrizing it using the function proposed by \citet{Bertola1991}:

\begin{equation}
    v(r) = v_{0} + \frac{v_{c}r}{(r^{2} + k^{2})^{\frac{\gamma}{2}}},
    \label{eq:RotCurvParam}
\end{equation}
where $v_{0}$ is the systemic velocity, $v_{c}$ is a parameter governing the amplitude of the rotation curve with $k$ describing its sharpness and $\gamma$ allowing rising or falling curves (with $\gamma=1$ for a flat rotation curve). When the spatial coverage of MaNGA was insufficient to reach $V_{max}$ an extrapolation to the rotation curve was applied (see left panel of figure \ref{fig:VelmapandCurve}). 

\begin{table*}[]
    \centering
    \caption{Linear best-fit parameters to scaling relations with integrated kinematics.}
%    \begin{adjustbox}{width=\textwidth}
{\scriptsize
\begin{tabular*}{\textwidth}{p{3.2cm}|p{0.9cm}|p{0.6cm}p{1.3cm}p{1.5cm}|p{0.6cm}p{1.3cm}p{1.5cm}|p{0.6cm}p{1.3cm}p{1.5cm}}
    \hline
         & No. & \multicolumn{3}{c|}{$M_{\star}-V_{R_e}$} & \multicolumn{3}{c|}{$M_{\star}-\sigma_{\star_e}$} & \multicolumn{3}{c}{$M_{\star}-S_{0.5}$} \\
        & Galaxies & Scatter & $b\pm1\sigma$  & $a\pm1\sigma$ & Scatter & $b\pm1\sigma$  & $a\pm1\sigma$ & Scatter & $b\pm1\sigma$  & $a\pm1\sigma$\\
        \hline
        This Work & 2458 & 0.23 & $0.21\pm0.01$ & $-0.36\pm0.08$ & 0.14 & $0.32\pm0.01$ & $-1.31\pm0.14$ & 0.1 & $0.31\pm0.01$ & $-1.22\pm0.04$ \\
        \citet{Barat2019MNRAS} & 270 & --- & --- & --- & --- & --- & --- & 0.05 & $0.36\pm0.01$ & $-1.62\pm0.06$\\
        \citet{Aquino-Ortiz2018} & 278 & 0.20 & $0.16\pm0.02$ & $0.32\pm0.30$ & 0.16 & $0.31\pm0.03$ & $-1.37\pm0.25$ & 0.08 & $0.26\pm0.01$ & $-0.78\pm0.1$\\
        \citet{Cortese2014} & 105 & 0.26 & --- & --- & 0.16 & --- & --- & 0.1 & $0.33\pm0.01$ & $-1.41\pm0.08$\\
        \hline
    \end{tabular*}
    }
%    \end{adjustbox}
    \label{tab:FitScaRel}
    \raggedright \textbf{Note.} All scatters are estimated from the linear best-fit. We consider stellar mass, $M_{\star}$, as independent variable. $log(V_{R_e},\sigma_{\star_e},S_{0.5})\ =\ a\ +\ blog(M_{\star})$. $V_{R_e},\ \sigma_{\star_e}$ and $S_{0.5}$ are given in [km/s],\ $M_{\star}$ in $M_\odot$.
\end{table*}
\section{Scaling Relations.} \label{sec:Results}

In this Section we show the analyzed kinematic scaling relations for galaxies segregated by early and late types. We include the sample of AGN's extracted from the current MaNGA sample derived following the criteria presented in \citet{Sanchez2018AGN}. In addition, we include some reference relations found by previous studies and the best fitted relations for our data set.
Table \ref{tab:FitScaRel} summarizes the results of our orthogonal linear fit along the horizontal axis considering the stellar mass on the vertical axis as the independent variable. We use the routines presented by \citet{Akritas1996ApJ} to fit the data points. It includes the zero-point, slope, and the scatter around the best-fitted relation. Outliers along this study are defined as data points beyond $2\sigma$ with respect to the main relation.

\subsection{$Stellar\ mass$\ $vs.$\ $integrated\ kinematics.$}

In the left-hand panel of Figure \ref{fig:Int_kin} we show the $M_{\star}-V_{R_e}$ relation. This relation has a large scatter of $0.23\ dex$ in $log(V_{R_e})$. This scatter is similar to the ones found by \citet{Aquino-Ortiz2018} for CALIFA galaxies ($0.20\ dex$) and by \citet{Cortese2014} ($0.26\ dex$) for galaxies from the SAMI survey. The huge scatter in these three samples is mostly dominated by the contribution of ETGs, that are mostly slow rotators \citep[e.g.][]{Emsellem2007MNRAS,Graham2018MNRAS,Falcon-Barroso2019A&A}. As a reference, we include the derivation of the stellar TF relation as presented in \citet{Avila-Reese2008}\footnote{We have reduced the stellar mass in \citet{Avila-Reese2008} by $0.09\ dex$ in order to convert from diet-Salpeter to Chabrier  IMF} and \citet{Aquino-Ortiz2018} both using $V_{max}$ for the rotation velocity. As expected, there is an offset between this classical TF derivation and our results at $R_{e}$. The scatter is also expected to be larger for the TF using $V_{R_e}$ than using $V_{max}$.

\begin{table*}[]
    \centering
        \caption{Linear fit parameters for the scaling relations with spatially resolved kinematics.}
    \footnotesize{
    \begin{tabular*}{\textwidth}{c|c|ccc|ccc}
    \hline
     \multicolumn{2}{c}{ Scaling relation} & \multicolumn{3}{|c|}{Tully-Fisher} & \multicolumn{3}{c}{$M_{\star}\ vs.\ S_{0.5}$}  \\
     \hline
     Authors & No. of Galaxies & Scatter (dex) & $b\pm1\sigma$ & $a\pm1\sigma$ & Scatter (dex) & $b\pm1\sigma$ & $a\pm1\sigma$ \\
     \hline
     This Work & 200 & 0.061 & $0.31\pm0.01$ & $-1.17\pm0.19$ & 0.066 & $0.34\pm0.01$ & $-1.57\pm0.12$ \\
     \citet{Avila-Reese2008} & 76 & 0.045 & $0.27\pm0.01$ & $-0.65\pm0.12$ & --- & --- & --- \\
     \citet{Aquino-Ortiz2018} & 92 & 0.053 & $0.30\pm0.02$ & $-1.00\pm0.02$ & 0.052 & $0.28\pm0.02$ & $-0.92\pm0.21$ \\
     \citet{Ferrero2017} & 7482 & 0.040 & $0.30\pm0.05$ & $-0.86\pm0.02$ & --- & --- & --- \\
     \hline
    \end{tabular*}
    }
    \label{tab:FitTFR}
    \raggedright \textbf{Note.} All scatters are estimated from the linear fit, we consider stellar mass, $M_{\star}$, as independent variable. $log(V_{max},S_{0.5})\ =\ a\ +\ blog(M_{\star})$. $V_{max}$ and $S_{0.5}$ are given in [km/s],\ $M_{\star}$ in $M_\odot$.
\end{table*}

In the central panel of Figure \ref{fig:Int_kin} we show the $M_{\star}-\sigma_{\star_e}$ relation. As a reference the FJ relation derived by \citet{Gallazzi2006} has been included for comparison. We find a scatter of $0.14\ dex$ in $log(\sigma
_{\star_e})$ with respect to the best fitted relation in our data, in agreement with the one found by \citet{Cortese2014} and \citet{Aquino-Ortiz2018} of $0.16\ dex$. However, the scatter for these three samples is larger than $0.07\ dex$, the one reported by \citet{Gallazzi2006} for only ETGs. As expected, ETGs follow the FJ relation. Contrary to the $M_{\star}-V_{R_e}$ relation where the scatter is dominated by ETGs, in the $M_{\star}-\sigma_{\star_e}$ relation the scatter is dominated by LTGs.

In the right-hand panel of Figure \ref{fig:Int_kin} we show the $M_{\star}-S_{0.5}$ distribution. The relations derived by \citet{Cortese2014} and \citet{Barat2019MNRAS} for the SAMI survey, as well as the one by \citet{Aquino-Ortiz2018} for the CALIFA data set have been included as a reference together with the best linear fit derived with our own MaNGA data. This $M_{\star}-S_{0.5}$ relation is clearly tighter than those relations using separately rotation or velocity dispersion. In this study for the MaNGA data set we find a scatter of $\sim 0.1\ dex$ in $log(S_{0.5})$. This scatter is in agreement with the ones reported for galaxies from the SAMI and CALIFA surveys (see Table \ref{tab:FitScaRel}). The reduction in the scatter when introducing the total velocity parameter, $S_{0.5}$, indicates that it is a better tracer to the circular velocity, i.e., the gravitational potential, than the rotation velocity and the velocity dispersion separately.

\subsection{Stellar mass vs. spatially resolved kinematics.}

\begin{figure}[ht!]
\centering
\includegraphics[width=0.5\textwidth, height = 8.5cm]{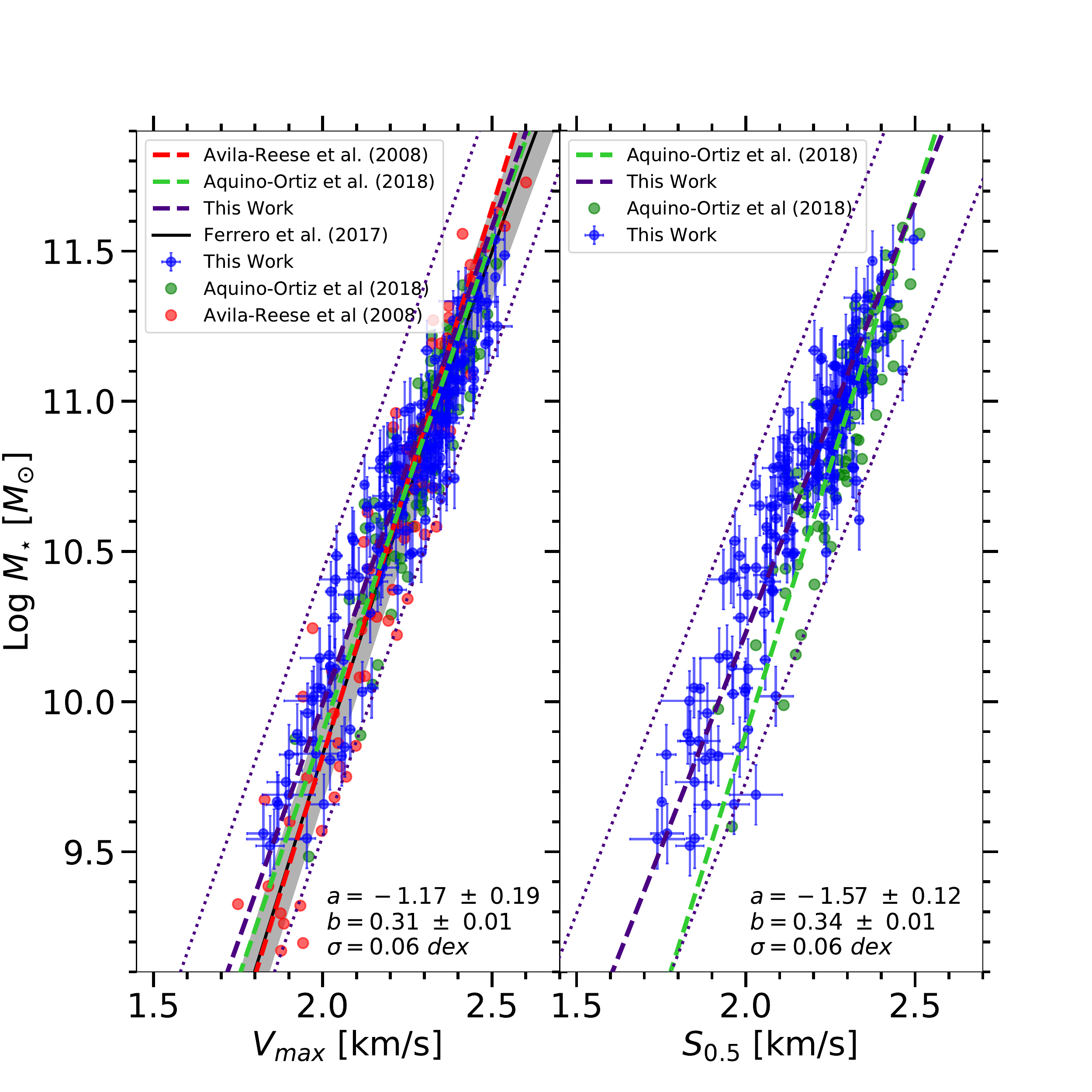}
\caption{Scaling relations with spatially resolved kinematics, $V_{max}$. \textit{Left panel:} Tully-Fisher relation; red symbols represents the data compilation from \citet{Avila-Reese2008}. The black solid line and the grey shaded band represent the predictions from \citet{Ferrero2017}. \textit{Right panel:} The $M_{\star}-S_{0.5}$ relation. In both panels blue and green symbols and dashed lines represents the results from the MaNGA sub-sample (used in this work) and the one published by \citet{Aquino-Ortiz2018} for the CALIFA survey.}
\label{fig:TFRVmax}
\end{figure}

Random and/or systematic errors could play an important role in the physical interpretation of scaling relations. They can modify the slope, zero-point and scatter. We have tried to narrow down their effects using our MaNGA sub-sample of good quality 200 LTGs described before. In particular, we tried to reproduce the ``Classical TF relation'' using the $V_{max}$ derived with the detailed spatially resolved kinematic analysis described in Section \ref{SRK}. In the left-hand panel of Figure \ref{fig:TFRVmax} we show our best TF relation compared with: (i) the one from \citet{Aquino-Ortiz2018} using a sub-sample of 92 spiral galaxies from the CALIFA survey; (ii) the relation from \citet{Avila-Reese2008} who used a compiled and homogenized sample; and (iii) the prediction from \citet{Ferrero2017} who used 7482 simulated galaxies at $z=0$ from the Evolution and Assembly of GaLaxies and their Environments (EAGLE) project \citep{Schaye2015MNRAS}. The best fitted parameters derived for the TF relation for these four samples, included the analyzed in this work, match pretty well, as shown in Table \ref{tab:FitTFR}.

In the right-hand panel of Figure \ref{fig:TFRVmax} we show the most precise estimation of the $M_{\star}-S_{0.5}$ relationship using $V_{max}$. We find differences in the slope and zero-point with respect to the sub-sample of 92 galaxies from the CALIFA survey published by \citet{Aquino-Ortiz2018} (See Table \ref{tab:FitTFR}). These variations may be induced due to differences in sample selection, survey systematics, and instrumental resolutions. Then it is hard to interpret the observed disagreement in slope and zero-points as physical differences. Despite of these discrepancies in the actual reported values, the trends that they trace are very similar, as seen in Figure \ref{fig:TFRVmax}.

The interesting result is that the ``Classical TF relation'' and the $M_{\star}-S_{0.5}$ using $V_{max}$ are tight, with the scatter very similar for the same sub-sample for disk-rotational-dominated systems (See Table \ref{tab:FitTFR}). In other words, if we include in both relations ETGs, which are velocity dispersion dominated systems, the scatter on the TF increases but does no significantly increase on the $M_{\star}-S_{0.5}$ (See Table \ref{tab:FitScaRel}). The total velocity parameter, $S_{0.5}$, which combine
rotation velocity and velocity dispersion seems to be the best tracer of the circular velocity reducing the scatter in the stellar mass-velocity relations. The same result was found by \citet{DeRossi2012} for simulations and by \citet{Aquino-Ortiz2018} for observational data from the CALIFA survey. 

\subsection{The Universal Fundamental Plane.}
\label{GP}

The so-called Universal Fundamental Plane is derived starting from a theoretical approach using the tensor virial theorem:

\begin{equation}
    \frac{1}{2}\frac{d^{2}\textbf{I}_{jk}}{dt^{2}} = 2\textbf{T}_{jk} + \textbf{$\Pi$}_{jk} + \textbf{W}_{jk},
    \label{eq.TVT}
\end{equation}
were \textbf{I} is the moment of inertia tensor, \textbf{T} and \textbf{$\Pi$} are the contributions of ordered and random motions to the kinetic energy tensor, respectively, and \textbf{W} is the potencial energy tensor. To rewrite Eq. (\ref{eq.TVT}) in terms of observed properties of galaxies, several simplifications and assumptions should be considered. In this study we use a purely empirical treatment and just enumerate and summarize them \citep[for further detailes see][]{Zaritsky2008,Zaritsky2012}: 

\begin{enumerate}
    \item Galaxies are in a steady state and the virial theorem holds over the effective radius. With this assumption the left-hand side of Eq. (\ref{eq.TVT}) is zero. To satisfy it, we discard merger and perturbed galaxies. We evaluate the trace of the resulting right-hand side of Eq. (\ref{eq.TVT}) and define the ordered and random contributions to the kinetic energy as $(1/2)A_{0}MV_{rot}^{2}$ and $A_{1}M\sigma_{\star}^{2}$, respectively. The potential energy is defined as $-B_{0}GM^{2}/R$. Hereafter, we define the characteristic radius, $R$, to be the effective radius $R_{e}$. Hence, Eq. (\ref{eq.TVT}) becomes:

    \begin{equation}
        A_{0}V_{R_e}^{2} + A_{1}\sigma_{\star_e}^{2} = B_{0}\frac{GM_{dyn_e}}{R_e},
    \label{eq:EVT}
    \end{equation}
    with $V_{R_e}$ as the stellar rotation velocity, $\sigma_{\star_e}$ the stellar velocity dispersion, $M_{dyn_e}$ the total dynamical mass enclosed at $R_e$, $G$ the gravitational constant, $A_{0},A_{1}$, and $B_{0}$ are correction factors obtained by fully evaluate the tensors. These correction factors could be different for each galaxy and also strong function of the formation history, dynamical state and environment of galaxies.
    \item The kinematic simplification. This means that galaxies are assumed to be isothermal spheres with isotropic velocity dispersion. Dividing the Eq. (\ref{eq:EVT}) by $A_{1}$, allow us to define the left-hand side as the total velocity parameter, $S_{K}^{2} = KV_{R_e}^{2}+\sigma_{\star_e}^{2}$
    \item The mass simplification. This means replace the total dynamical mass at the effective radius, $M_{dyn_e}$, with observable properties like the dynamical mass-to-light ratio within $R_{e}$, $\Upsilon_e$, and the luminosity, $L_{e}$, i.e., $M_{dyn_e}=\Upsilon_{e}L_{e}$. Thus, it is assumed that  $\Upsilon_{e}$ is constant within the considered aperture.
    \item Homology, which implies that galaxies live on a plane in the $(R_e, I_e, S_K)$ space. Thus, the correction factors $A_{0}, A_{1}$ and $B_{0}$ are very similar among galaxies.
\end{enumerate}

Applying the previous simplifications and assumptions we can rewrite the tensor virial theorem in terms of observational properties as follows:

\begin{equation}
    S_{K}^{2} = B_{0}A_{1}^{'}G\pi \Upsilon_{e} R_{e}I_{e}.
\end{equation}
As indicated before, it is found that the minimum scatter in the $M_{\star}-S_{K}$ relation is achieved when $K=0.5$ \citep[e.g.][]{Cortese2014, Aquino-Ortiz2018, Gilhuly2019,Barat2019MNRAS}. Therefore, from now on, we will fix $K=0.5$ on the total velocity parameter. Finally, we define a normalization constant $C=G\pi B_{0}A_{1}^{'}$ and take the logarithm to define the Universal Fundamental Plane as follows: 

\begin{figure}[ht!]
\centering
\includegraphics[width=0.5\textwidth, height=10cm]{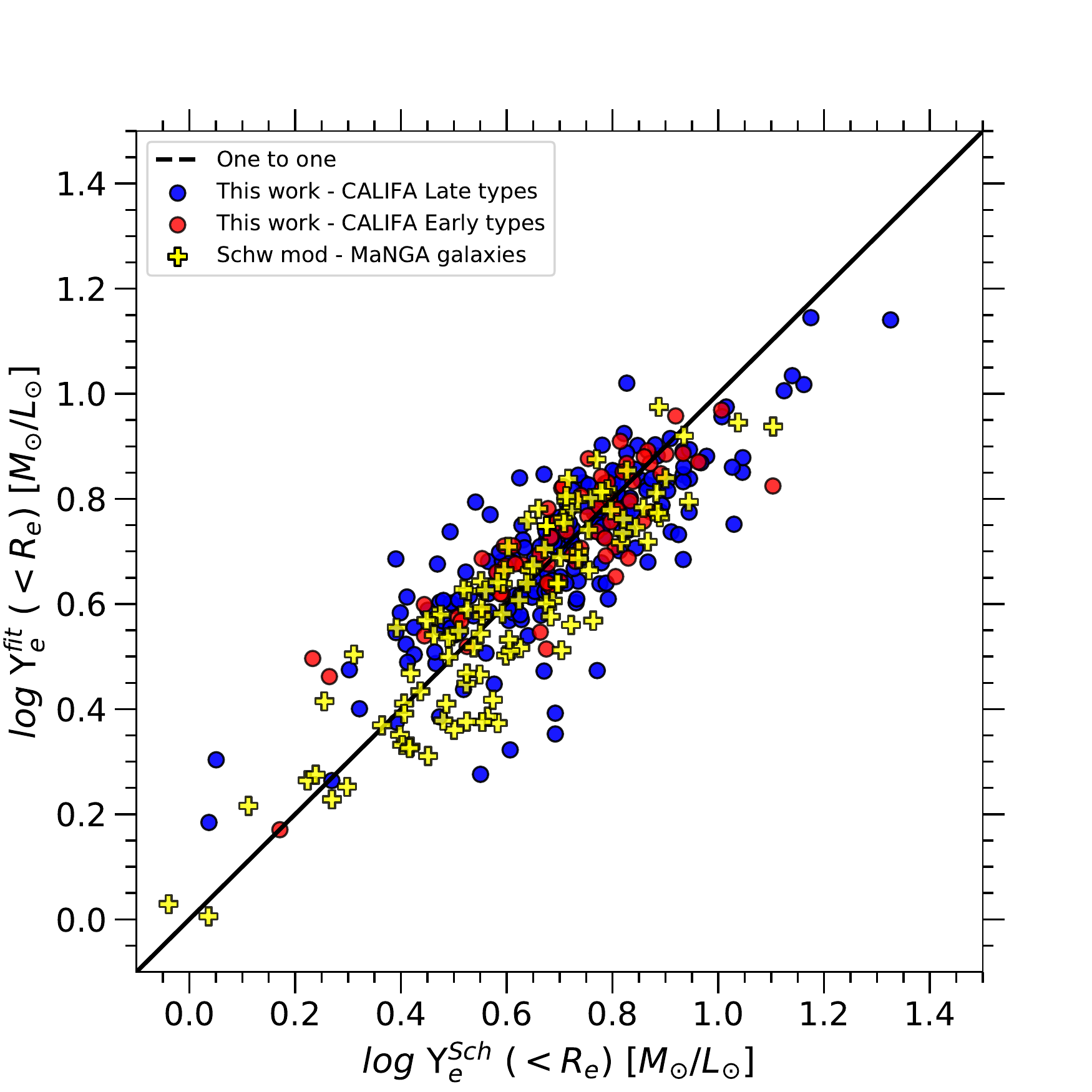}
\caption{Comparison between the dynamical mass-to-light ratios derived from robust Schwarzschild dynamical models, $\Upsilon_{e}^{Sch}$, and our estimates using Eq. \ref{eq:MLR} and the $\beta_{i}'s$ calibrated coefficients, $\Upsilon_{e}^{fit}$. Red (Blue) symbols are early (late) type galaxies from the CALIFA sub-sample by \citet{Zhu2018}. The yellow symbols are the values for the MaNGA galaxies analyzed by \citet{Yunpeng2020}.} 
\label{fig:CALIFA_MLfitobs}
\end{figure}

\begin{equation}
log(\Upsilon_{e}) = log (S_{0.5}^{2}) - log (I_{e}) - log (R_{e}) + C.
\label{eq:GFP}
\end{equation}

\begin{figure*}[ht!]
\centering
\includegraphics[width=0.85\textwidth, height=12cm]{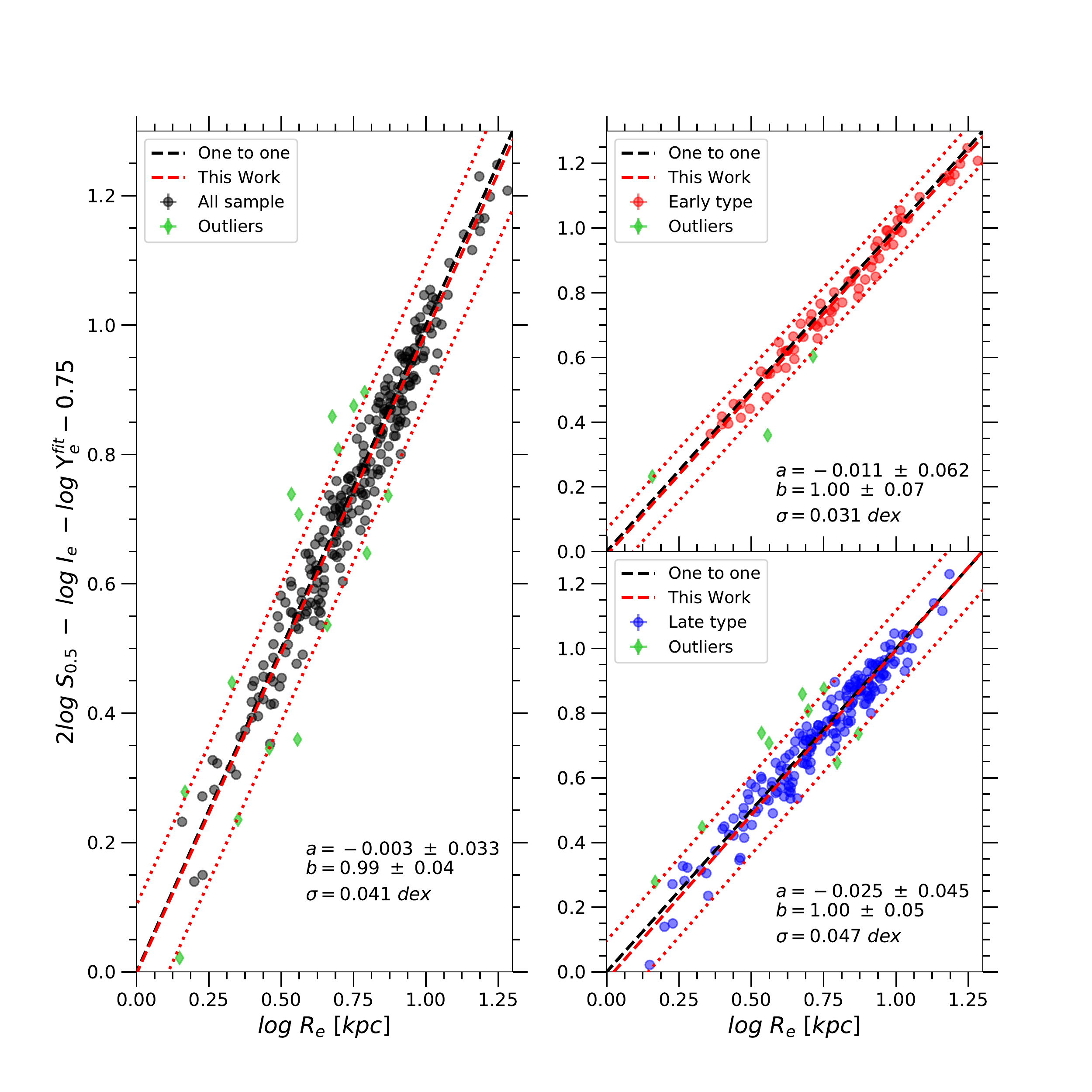}
\caption{The CALIFA Universal Fundamental Plane. \textit{Left panel:} the full sub-sample of 300 galaxies. The dashed black and red lines represents the one-to-one relationship and the best-fit to the data points, respectively. \textit{Top right-hand panel:} early type galaxies. \textit{Bottom right-hand panel:} late type galaxies. Red dotted lines in all panels marks the $2\sigma$ bands. The green symbols are the outliers.}
\label{fig:CALIFA_GP}
\end{figure*}

Since \citet{Zaritsky2008,Zaritsky2012} had no information for the $\Upsilon_{e}$ for their full sample, they proposed a fitting function that depend on distance independent variables, $S_{0.5}$, and $I_{e}$, including second-order and cross-terms to to study the General Fundamental Manifold. In Appendix \ref{apx:FM}, we explore it following this approach, for the MaNGA data set, for which we do not have either an independent estimation of $\Upsilon_e$.
However, in this study we follow the suggestion by \citet{Zaritsky2008,Zaritsky2012} to use a sample of galaxies with independent estimations for the dynamical mass-to-light ratio, $\Upsilon_{e}$, to calibrate the UFP. To do this, we use the subset of 300 CALIFA galaxies presented in Section \ref{sec:CALIFA_DATA}. For this subset, the dynamical mass-to-light ratio, $\Upsilon_{e}^{Sch}$, has been constrained accurately by means of a dynamical analysis based on the Schwarzschild orbit super-position method by \citet{Zhu2018}. Once it has been calibrated, one can use it to solve for the $log\ (\Upsilon_{e})$ for any galaxy with measured $S_{0.5}$, $I_e$, and $R_e$.

For the calibration, we apply a multiple-linear regression to recover the dynamically estimated $log(\Upsilon_{e}^{Sch})$ (defined as the dependent variable), in terms of the observed data $S_{0.5}, I_{e}$, and $R_{e}$ (independent variables) as follow:

\begin{equation}
    log(\Upsilon_{e}^{Sch}) = \beta_{0}\ +\ \beta_{1}log(S_{0.5})\ +\ \beta_{2}log(I_{e})\ +\ \beta_{3}log(R_{e}),
    \label{eq:MLR}
\end{equation}

\begin{figure*}[ht!]
\centering
\includegraphics[width=0.85\textwidth, height=12cm]{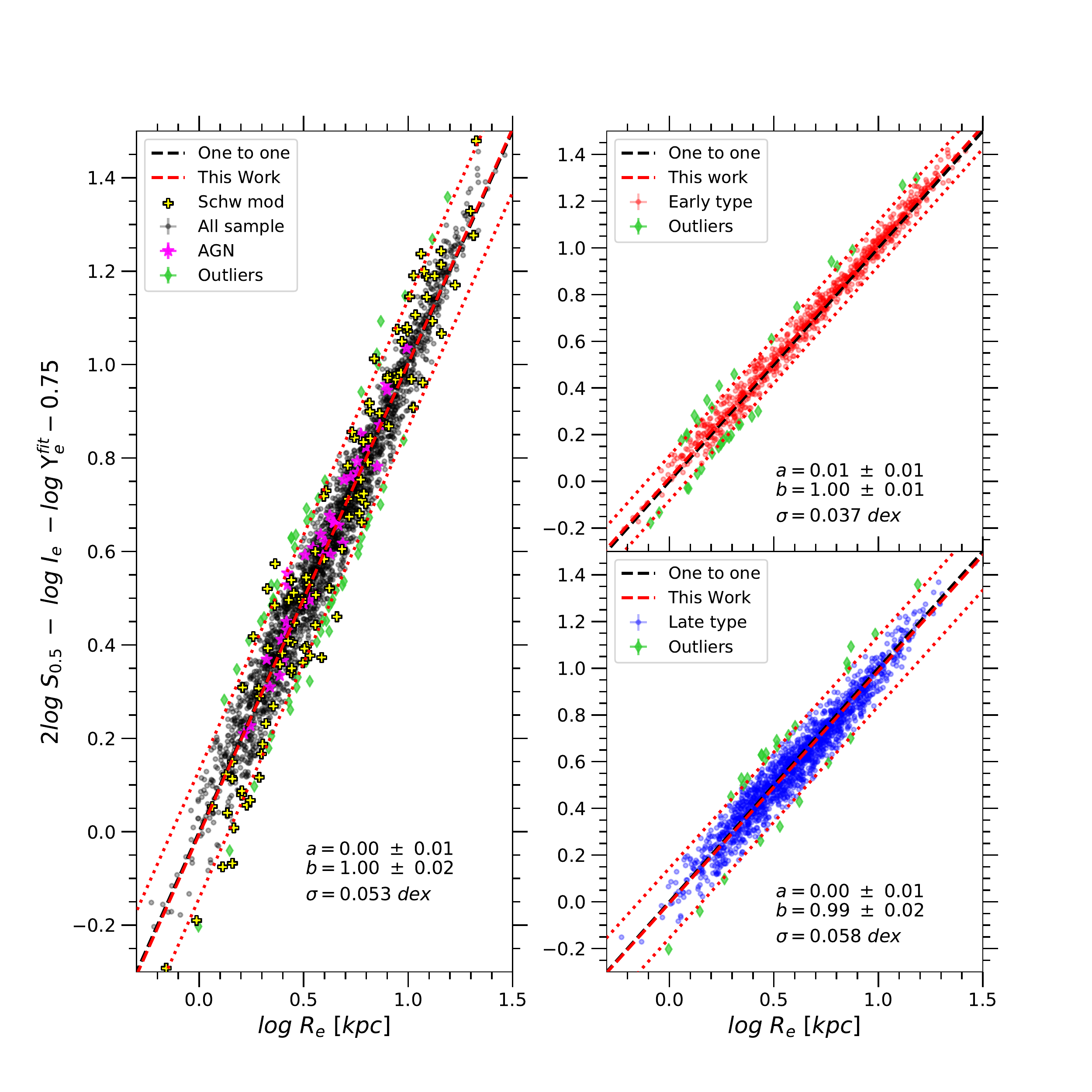}
\caption{The MaNGA Universal Fundamental Plane. \textit{Left panel:} black symbols represent the sample of $2458$ galaxies of this study, magenta symbols are the sample of AGN's from \citet{Sanchez2018AGN} and the yellow ones represents the 108 galaxies from the MaNGA sample with independent estimations of the dynamical mass-to-light ratio, $\Upsilon_{e}^{Sch}$, by \citet{Yunpeng2020}. \textit{Top right-hand panel:} early type galaxies. \textit{Bottom right-hand panel:} late type galaxies. In all panels: the green symbols are the outliers. The dashed black and red lines represents the one-to-one relationship and the best-fit to the data points. The dotted red lines marks the $2\sigma$ band.}
\label{fig:MaNGA_GP}
\end{figure*}

were the $\beta_{i}$ are the coefficients for each independent variable. The calibration that best recover the $log(\Upsilon_{e}^{Sch})$ yields the following values: $\beta_{0}=-0.53\pm0.1$, $\beta_{1}=1.49\pm0.08$, $\beta_{2}=-0.72\pm0.03$ and $\beta_{3}=-0.63\pm0.05$. We use these coefficients together with the 3 independent variables $(S_{0.5}, I_{e}, R_{e})$ to calculate the fitted dynamical mass-to-light ratios, $log\ (\Upsilon_{e}^{fit})$. The obligatory exercise is to compare the dynamically-determined values, $log(\Upsilon_{e}^{Sch})$, with the ones derived from our calibration, $log(\Upsilon_{e}^{fit})$. We present the results of this exercise in Figure \ref{fig:CALIFA_MLfitobs}, where it is shown both parameters for the CALIFA analyzed subset of galaxies. In addition, we include the recent estimations for the $log(\Upsilon_{e}^{Sch})$ presented by \citet{Yunpeng2020} on 108 ETGs extracted from the MaNGA survey compared with our estimated $log(\Upsilon_{e}^{fit})$. The comparison for both, CALIFA and MaNGA data, follows the one-to-one relationship with a scatter of $0.09\ dex$ for LTGs and $0.07\ dex$ for ETGs in good agreement with the reported value of $0.06\ dex$ by \citet{Zaritsky2008}, who use a set of nearby spheroidal galaxies.
In Figure \ref{fig:CALIFA_GP} we show the UFP for the CALIFA sub-sample of 300 galaxies as a result of replacing the $log(\Upsilon_{e})$ by the fitted $log(\Upsilon^{fit}_{e})$ and rearrange terms in Equation (\ref{eq:GFP}). By construction the defined UFP should be a one-to-one relationship. This plane shows a scatter of $\sim\ 0.04\ dex$ in $log(R_e)$ with $\sim 5\%$ of outliers (15 galaxies) and an average value for the coefficient $C=-0.75$, in good agreement with \citet{Zaritsky2008,Zaritsky2012}. The low scatter of this one-to-one relationship suggests that the value of the coefficient $(C=G\pi B_{0}A_{1}^{'})$ presents a very narrow range of variation among different galaxies. 

For the MaNGA data set explored in this study, we use the effective radius $R_{e}$, and surface brightness $I_{e}$, described in the Section \ref{sec:MaNGA_DATA}. The total velocity parameter $S_{0.5}^{2} =0.5V_{R_e}+\sigma_{\star_e}$, is calculated using the stellar mean velocity $V_{R_e}$, and stellar velocity dispersion $\sigma_{\star_e}$ estimated using the analysis presented in the Section \ref{subsec:Int_kin} and plotted in the right-hand side of Figure \ref{fig:Int_kin}. Nevertheless, we do not have independent estimates for the dynamical mass-to-light ratios $\Upsilon_{e}$ for our full MaNGA sample. Instead of that, we used the fitted ones, $\Upsilon_{e}^{fit}$, estimated with the Equation (\ref{eq:MLR}) and the $\beta_{i}'s\ $ coefficients calibrated previously with the CALIFA sample.
In Figure \ref{fig:MaNGA_GP} we show the derived UFP for the MaNGA data set. For comparison purposes we use the data published by \citet{Yunpeng2020}, already presented in Figure \ref{fig:CALIFA_MLfitobs}. The relation for the full sample and independent estimations of the $\Upsilon_{e}^{Sch}$ follows, as expected, the one-to-one relationship with $\sim 3\%$ of outliers (60 galaxies) and a scatter of $0.05\ dex$ slightly larger that the one found previously using the CALIFA sub-sample ($0.04\ dex$) but in good agreement with the reported by \citet{Zaritsky2008} ($\sim0.05\ dex$) for their sub-sample with independent estimations of the $\Upsilon_e$. Am interesting results here is that the sample of AGN's follows the one-to-one relationship, none of them is an outlier. This result suggests that the role of the AGN in the galactic dynamics and structure is unimportant, at least at the effective radius.

\section{The Dynamical mass estimator}
\label{sec:mass_estimator}

Historically, the virial theorem has been the primary tool to determine the dynamical mass of galaxies \citep[e.g.][]{Zwicky1933}. Hence, the $\Upsilon_{e}^{fit}$, derived from the tensor virial theorem can be used to measure dynamical masses at $R_e$. We apply the Eq. (\ref{eq:MLR}) and the $\beta_{i}'s$ coefficients to get $\Upsilon_{e}^{fit}$, then we multiplying by the luminosity, $L_{e}$, to estimate the dynamical mass. 

\begin{figure}[ht!]
\centering
\includegraphics[width=0.5\textwidth,height=9cm]{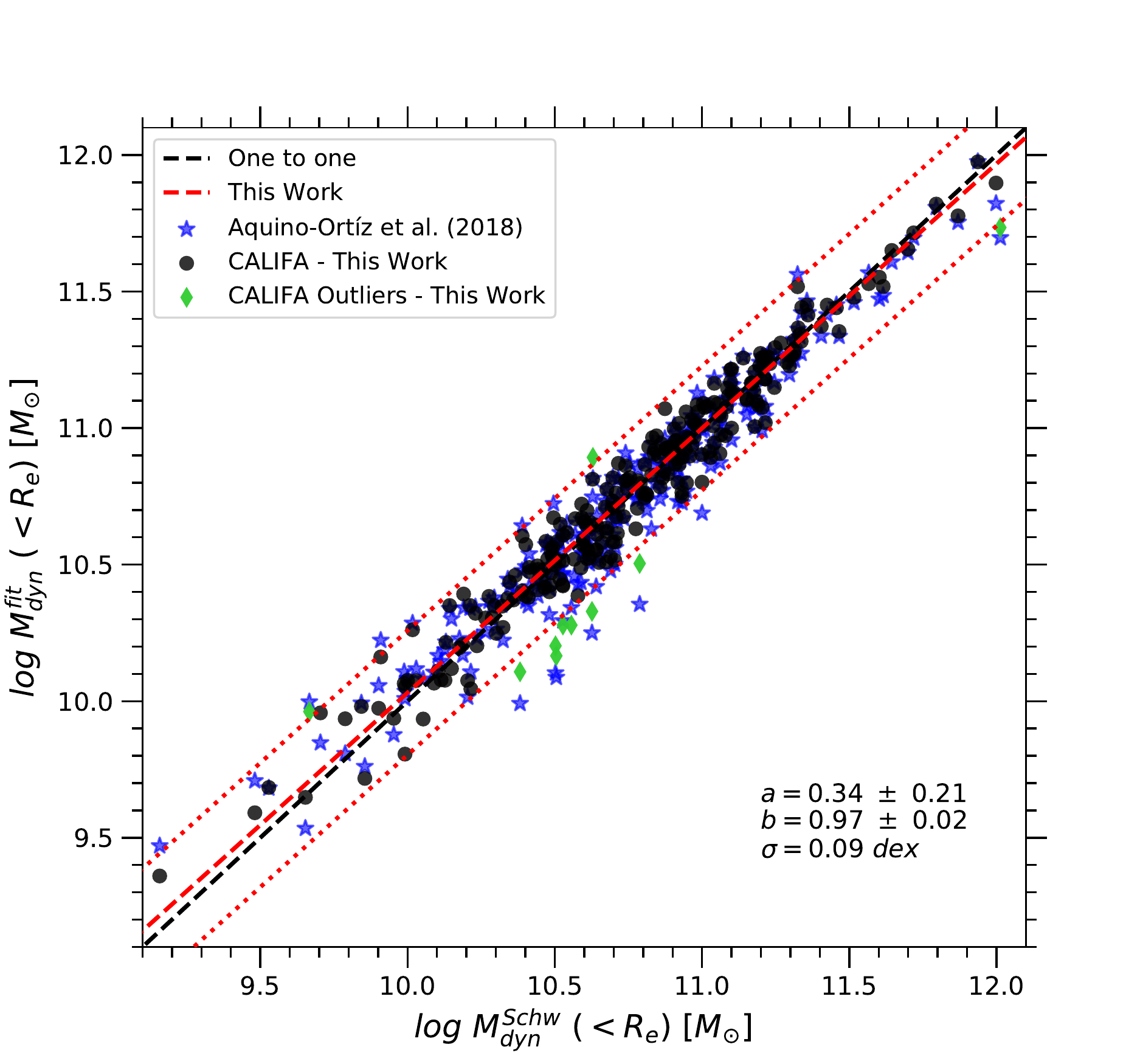}
\caption{One-to-one relationship between dynamical masses from the sub-sample of 300 CALIFA galaxies. Black symbols represents the comparison between the dynamical masses inferred from dynamical models by \citet{Zhu2018} and the ones form this analysis using the kinematics and surface brightness (Eq. (\ref{eq:MLR})). The $M_{dyn}^{fit}$ are estimated with the $\Upsilon_{e}^{fit}$ multiplied by the luminosity $L_{e}$. Blue symbols represent the comparison between the dynamically estimated and the ones using only the total velocity parameter $S_{0.5}$ by \citet{Aquino-Ortiz2018}. Green symbols are outliers.}
\label{fig:Dynmass_GP}
\end{figure}

In Figure \ref{fig:Dynmass_GP} we plot the one-to-one relation between the estimated dynamical mass derived adopting the described procedure, and the inferred from the dynamical modelling by \citet{Zhu2018} for the sub-sample of 300 CALIFA galaxies analyzed along this article. We also include our previous published estimations, based on just the kinematic parameter $S_{0.5}$  \citep[see Eq. (5) in][]{Aquino-Ortiz2018}. We find that our new estimations of the dynamical masses are consistent with those derived by dynamical models within a scatter of $0.09\ dex$. This scatter is smaller than the one of $0.15\ dex$ found using only the total velocity parameter $S_{0.5}$. This result is expected in the sense that the Eq. (\ref{eq:MLR}) includes more information of galaxies, combining the kinematics, luminosity and the scale-length (reinforcing the idea that all those parameters are indeed important to derive the dynamical stage of a galaxy). This is indeed the same information that the Schwarzschild orbit-superposition method uses to build dynamical models \citep[e.g.][]{Remco2008MNRAS}.

\subsection{The dynamical-to-stellar mass relation.}
\label{Mdyn-Mstar}

\begin{figure*}[ht!]
\centering
\includegraphics[width=0.9\textwidth,height=13cm]{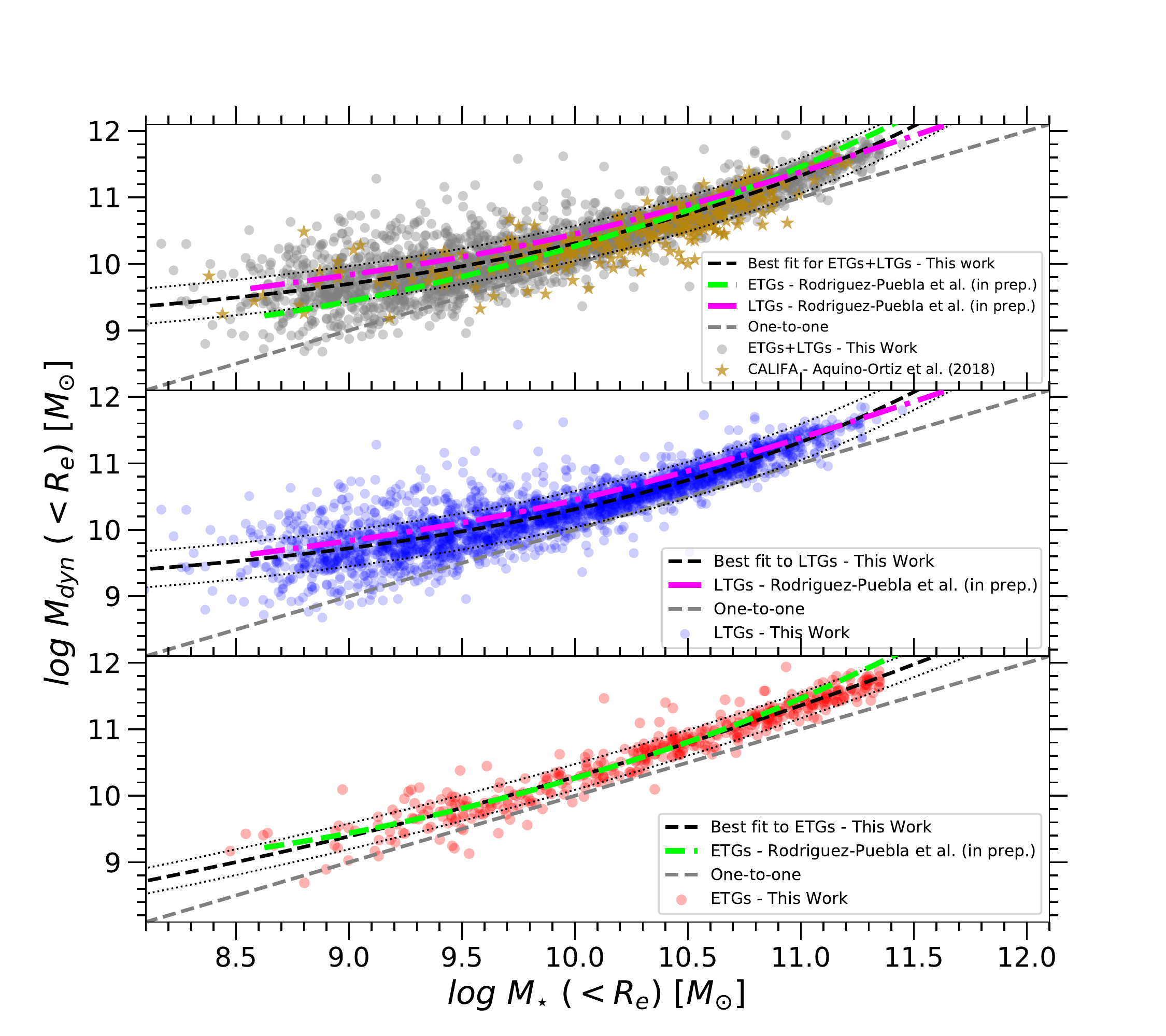}
\caption{The $M_{dyn}-M_{\star}$ relation at $R_e$. \textit{Top panel:} grey symbols represents the relation for the full MaNGA sample, and orange ones for galaxies from the CALIFA survey by \citet{Aquino-Ortiz2018}. \textit{Middle panel:} The relation for LTGs. \textit{Bottom panel:} The distribution for ETGs. In all panels green and magenta dashed lines represents the predictions for ETGs and LTGs from semi-empirical models by Rodriguez-Puebla et al. (in prep). Black dashed and dotted lines are the best-fit and the $1\sigma$ bands for the observed distribution. Grey dashed line is the one-to-one comparison.} 
\label{fig:MaNGA_Mdyn}
\end{figure*}

For the MaNGA sample we do not have independent information about the dynamical masses to perform a direct comparison. Instead of that, following our previous explorations presented in \citet{Aquino-Ortiz2018} we study the distribution of stellar masses along the dynamical ones. The top panel of Figure \ref{fig:MaNGA_Mdyn} presents the $M_{dyn_e}-M_{\star_e}$ relation for the MaNGA data set, together with the results for galaxies from the CALIFA survey \citep{Aquino-Ortiz2018}. We find good agreement between the distributions for CALIFA and MaNGA galaxies. We characterize the observed distribution for the MaNGA data set with the following functional form:

\begin{equation}
    log(M_{dyn_e}) = \alpha_1\times \alpha_2^{log(M_{\star_e})} + \alpha_3.
\end{equation}

In Table \ref{tab:Mdyn_Mste} we report the best-fit parameters to the full sample, early and late galaxy types.

For comparison proposes, we also include predictions from a semi-empirical modelling approach by Rodriguez-Puebla et al. (in prep.). For the semi-empirical modelling the authors generated a complete population of galaxies by loading the bulge/disc systems into $\Lambda$ Cold Dark Matter haloes, taking into account the adiabatic contraction of the inner halo by the baryons, and following the semi-empirical stellar-to-halo mass relations (hereafter SHMR) of late- and early-type galaxies.

\begin{table}[h!]
    \centering
        \caption{Best-fit parameters to the $M_{dyn_e}-M_{\star_e}$ relation for the MaNGA data set.}
    \begin{tabular}{l|c|c|c|c}
    \hline
    Sample & $\alpha_1\ \pm 1\sigma$ & $\alpha_2\ \pm 1\sigma$ & $\alpha_3\ \pm 1\sigma$ & Scatter \\
    \hline
    Full & 0.009 $\pm$ 0.003 & 1.676 $\pm$ 0.045 & 8.79 $\pm$ 0.08 & 0.26\\
    LTGs & 0.006 $\pm$ 0.003 & 1.719 $\pm$ 0.057 & 8.90
    $\pm$ 0.09 & 0.27\\
    ETGs & 0.811 $\pm$ 0.521 & 1.205 $\pm$ 0.049 & 5.03
    $\pm$ 1.20 & 0.19\\
    \hline
    \end{tabular}
    \label{tab:Mdyn_Mste}
\end{table}

The medium (bottom) panel of Figure \ref{fig:MaNGA_Mdyn} shows the $M_{dyn}-M_{\star}$ relation for LTGs (ETGs) and a comparison between the best-fit estimation provided by our analysis and the semi-empirical prediction. We find a remarkable agreement with the theoretical approach. For LTGs, below $\sim 8\times 10^{10} M_\odot$ there is a clear deviation, i.e., galaxies show larger dynamical masses than their stellar ones, which indicates that in the inner regions of galaxies at low-mass regime appear to be more dark matter dominated as less massive they are. Whereas for more massive galaxies the deviation is weaker. For ETGs the distribution follows a nearly linear relation at all masses with a slight bend for low-mass galaxies. 
Both the observed distribution for the MaNGA data set and the predictions for the semi-empirical models follow similar trends. The bends seen in the predictions are well understood. As we mentioned before, by construction, the semi-empirical models follows the SHMR, which bends at lower $M_{\star}/M_{vir}$ ratios both at lower and higher masses \citep[e.g. See the bottom panel of Figure 5 in  ][]{Rodriguez-Puebla2015ApJ}. The shape of this SHMR is inherited to the predicted inner mass distribution, hence to the $M_{\star}-M_{dyn}$ relation at $R_e$. Therefore, our observed distribution, which agree with the predicted one, could be an important tool to attain a connection between the inner galaxy dynamics of the local galaxy population and the properties of the cosmological dark matter haloes.
\section{Discussion.}
\label{Discussion}

We confirm that all galaxies from the MaNGA sample regardless of the early/late morphological type lie into the same $M_{\star}-S_{0.5}$ relationship. The scatter of $0.1\ dex$ is in agreement with previous studies using different surveys, such as SAMI \citep{Cortese2014,Barat2019MNRAS} and CALIFA \citep{Aquino-Ortiz2018,Gilhuly2019}. The remarkable reduction of the scatter points towards a more complex internal kinematics in galaxies. LTGs, though are rotation-dominated systems, frequently show no-circular(random) motions \citep[e.g.][]{Zhu2018b,Cortese2014,Aquino-Ortiz2018}. On the other hand, ETGs, velocity dispersion-dominated systems, sometimes present a fraction of rotation \citep[e.g.][]{Emsellem2007MNRAS,Cappellari2011MNRAS,Graham2018MNRAS,Falcon-Barroso2019A&A}. The combination of $V_{R_e}$ and $\sigma_{\star_e}$ in a single parameter $S_{0.5}$, provides a better proxy for the circular velocity (i.e., the gravitational potential or dynamical mass) of a galaxy \citep[e.g.][]{Aquino-Ortiz2018}.

All galaxy types within the analyzed CALIFA sub-sample of 300 galaxies with independent estimations of the dynamical mass-to-light ratio, $\Upsilon_e^{Sch}$, and for the MaNGA data set using the calibrated $\Upsilon_{e}^{fit}$, fall on the so-called as Universal Fundamental Plane. The low scatter observed in the full CALIFA sub-sample ($\sim 0.04\ dex$) and MaNGA sample ($\sim 0.05\ dex$) about the mean relation is comparable or even lower to the ones observed in the TF relation for spiral galaxies \citep[e.g.][$\sim 0.05\ dex$]{Avila-Reese2008,Aquino-Ortiz2018}, and the Fundamental Plane for early type galaxies \citep[e.g.][$\sim 0.09\ dex$]{Cappellari2013}. Furthermore, the scatter for ETGs ($0.037\ dex$) and LTGs ($0.058\ dex$) in the UFP for the MaNGA sample are lower than the ones reported by \citet[][]{Li2018MNRAS} in the Mass Plane ($0.047\ dex$ and $0.061\ dex$ for early and late galaxy types, respectively). This reduced scatter is because our approach is more general including more information of galaxies, such as the surface brightness, $I_e$, and dynamical mass-to-light ratio, $\Upsilon_e$.
\citet{Zaritsky2008} claims that the approach used to explore the UFP could fail for gas rich galaxies. In other words, for systems where the majority of the baryons are in the gas instead of the stars, then the ratio between the dynamical mass and the optical luminosity from stars $\Upsilon_{e}=M_{dyn_e}/L_{e}$, become larger. In our analysis the scatter for LTGs is slightly bigger than the one for ETGs in all the relations explored along this paper (see Figures \ref{fig:Int_kin}, \ref{fig:CALIFA_MLfitobs}, \ref{fig:CALIFA_GP}, \ref{fig:MaNGA_GP} and \ref{fig:MaNGA_Mdyn}). This is because measure the kinematic parameters, effective radius, and dynamical masses on galaxies with emission lines, younger and more metal poor as the LTGs are, is less accurate due to the limited spectral resolution.

\citet{Zaritsky2008} suggest that the origin of this small scatter could be; (i) by internal factors, such as stellar orbital structure, nuclear activity, (ii) by mass loss history, and (iii) by external factors, such as environment or accretion history. However, we find that the nuclear activity does not contribute to the scatter. The sample of AGN's follows the one-to-one relationship, none of them is outlier. The nature of the 3-to-5\% of outlier galaxies found in Figures \ref{fig:CALIFA_GP} and \ref{fig:MaNGA_GP} deviating from the main trend could provide key information about the nature of the scatter. Those galaxies will be explored in a forthcoming article using state-of-the-art dynamical orbital modellings, following \citet{Zhu2018} and \citet[][]{Yunpeng2020}.
\section{Summary and conclusions}
\label{Summary}

Using $2458$ galaxies observed with integral field spectroscopy from the MaNGA survey, we re-examine the $M_{\star}-S_{0.5}$ scaling relation. We also study a Universal Fundamental Plane for early and late galaxy types calibrated with a sub-sample of 300 galaxies from the CALIFA survey adopting a totally empirical approach.
We summarize the main results of this study as follows:

(i) We confirm that early and late galaxy types together follow the $M_{\star}-S_{0.5}$ relationship with a remarkable reduction of scatter compared to the individual $M_{\star}-V_{R_e}$ and $M_{\star} - \sigma_{\star_e}$ relations. The scatter on both later relations is dominated by early- and late-types, respectively, in agreement with previous studies using data from the SAMI  \citep[e.g.][]{Cortese2014,Barat2019MNRAS} and CALIFA \citep[e.g.][]{Aquino-Ortiz2018,Gilhuly2019} surveys.

(ii) We calibrate the Universal Fundamental Plane with a sub-sample of 300 galaxies from the CALIFA survey with independent estimations of the dynamical mass-to-light ratios at $R_e$, $\Upsilon_{e}^{Sch}$, surface brightness, $I_{e}$, and total velocity parameter, $S_{0.5}$. We find that all classes of galaxies, from spheroids to disks, follow this Universal Fundamental Plane with a scatter significantly smaller than the one reported for the $M_{\star}-S_{0.5}$ relation. The scatter about that surface $(\sim 0.04\ dex)$ is comparable (or smaller) than the ones observed in Tully-Fisher, $\sim 0.05\ dex$ (Fundamental Plane, $\sim 0.09\ dex)$ studies but for a wider range of galaxy types.

(iii) We propose a simple but competitive procedure to estimate the dynamical mass-to-light ratio, $\Upsilon_{e}$, in galaxies (hence the dynamical mass) at $R_{e}$, easier to apply to massive surveys than more detailed analysis, although with lower precision.

(iv) We use the estimated dynamical mass-to-light ratio, $\Upsilon_{e}^{fit}$, from our analysis to explore the Universal Fundamental Plane with the MaNGA data set. The results are consistent with the ones from the CALIFA sub-sample with a slightly larger scatter ($\sim 0.05\ dex$) in good agreement with the one of $\sim 0.05\ dex$ suggested by \citet{Zaritsky2008}

(v) We show that AGN hosts follow the same one-to-one relationship within $1\sigma$ as the general population of galaxies. Furthermore, none of them is an outlier. This result could suggest that the role of the nuclear activity, is unimportant in determining the inner structure of galaxies.

(vi) We find $\sim 3\%$ of outliers in the Universal Fundamental Plane for both the CALIFA and MaNGA sub-samples. They could provide key information about the scatter. We will explore the nature of those outliers in an upcoming dynamical study.

(vii) Finally, we find a remarkable agreement between the observed $M_{dyn_e}-M_{\star_e}$ distribution and the predicted with semi-empirical modelling approach. This relation could be a projection of the SHMR at the inner part of galaxies, therefore a connection between the inner galaxy dynamics of the local population of galaxies and the properties of the $\Lambda CDM$ haloes.
\section{A C K N O W L E D G E M E N T S}

We are grateful for the support of a CONACYT grant CB-285080 and FC-2016-01-1916, and funding from the PAPIIT-DGAPA-IN100519 (UNAM) project. O.V. and E.A. acknowledge support from PAPIIT-DGAPA: IN112518 and IG101620 UNAM grants. 
%S.F.S. thank CONACYT grant CB285080
%and funding from the PAPIIT-DGAPA-IA101217 (UNAM) project.
JKBB thanks funding from the PAPIIT-DGAPA-IA100420 (UNAM) project. GvdV acknowledges funding from the European Research Council (ERC) under the European Union's Horizon 2020 research and innovation programme under grant agreement No 724857 (Consolidator Grant ArcheoDyn). This study makes use of data from the CALIFA (\url{https://califa.caha.es/}) and MaNGA (\url{https://www.sdss.org/surveys/manga/}) surveys. Data used in this research was supported through computational and human resources provided by the LAMOD UNAM project.LAMOD is a collaborative effort between the IA, ICN and IQ institutes at UNAM.\\ 
Funding for the Sloan Digital Sky Survey IV has been provided by the Alfred P. Sloan Foundation, the U.S. Department of Energy Office of Science, and the Participating Institutions. SDSS acknowledges support and resources from the Center for High-Performance Computing at the University of Utah. The SDSS web site is www.sdss.org.

SDSS is managed by the Astrophysical Research Consortium for the Participating Institutions of the SDSS Collaboration including the Brazilian Participation Group, the Carnegie Institution for Science, Carnegie Mellon University, the Chilean Participation Group, the French Participation Group, Harvard-Smithsonian Center for Astrophysics, Instituto de Astrofísica de Canarias, The Johns Hopkins University, Kavli Institute for the Physics and Mathematics of the Universe (IPMU) / University of Tokyo, the Korean Participation Group, Lawrence Berkeley National Laboratory, Leibniz Institut für Astrophysik Potsdam (AIP), Max-Planck-Institut für Astronomie (MPIA Heidelberg), Max-Planck-Institut für Astrophysik (MPA Garching), Max-Planck-Institut für Extraterrestrische Physik (MPE), National Astronomical Observatories of China, New Mexico State University, New York University, University of Notre Dame, Observatório Nacional / MCTI, The Ohio State University, Pennsylvania State University, Shanghai Astronomical Observatory, United Kingdom Participation Group, Universidad Nacional Autónoma de México, University of Arizona, University of Colorado Boulder, University of Oxford, University of Portsmouth, University of Utah, University of Virginia, University of Washington, University of Wisconsin, Vanderbilt University, and Yale University.

\twocolumngrid

\appendix

\section{Fundamental Manifold}
\label{apx:FM}

Initially, \citet{Zaritsky2008,Zaritsky2012} proposed the Fundamental Manifold as a result of a fitting function for $log\ \Upsilon_{e}$. Because they didn't have independent estimations for the $\Upsilon_{e}$ for its sample of 1925 galaxies, that fitting function depends only on two variables that are distance independent: the total velocity parameter $S_{0.5}$, and the i-band surface brightness $I_{e}$, including a second order and cross-terms as follows:

\begin{equation}
    log(\Upsilon_{e}^{fit}) = \alpha_{1}log(S_{0.5}) + \alpha_{2}log(I_{e}) + \alpha_{3}log^{2}(S_{0.5}) + \alpha_{4}log^{2}(I_{e}) + \alpha_{5}log(S_{0.5})log(I_{e})+\alpha_{6},
\label{eq:new}
\end{equation}

where the $\alpha_{i}$ are the coefficients of the adopted functional form. They replaced $log \Upsilon_{e}-C$ with $log \Upsilon_{e}^{fit}$ in Eq. \ref{eq:GFP} and plot a re-arrangement of the terms. They found that all classes of galaxies lie on the Fundamental Manifold with a scatter of $0.1\ dex$.

\begin{figure*}[ht!]
\centering
\includegraphics[width=0.9\textwidth, height=12cm]{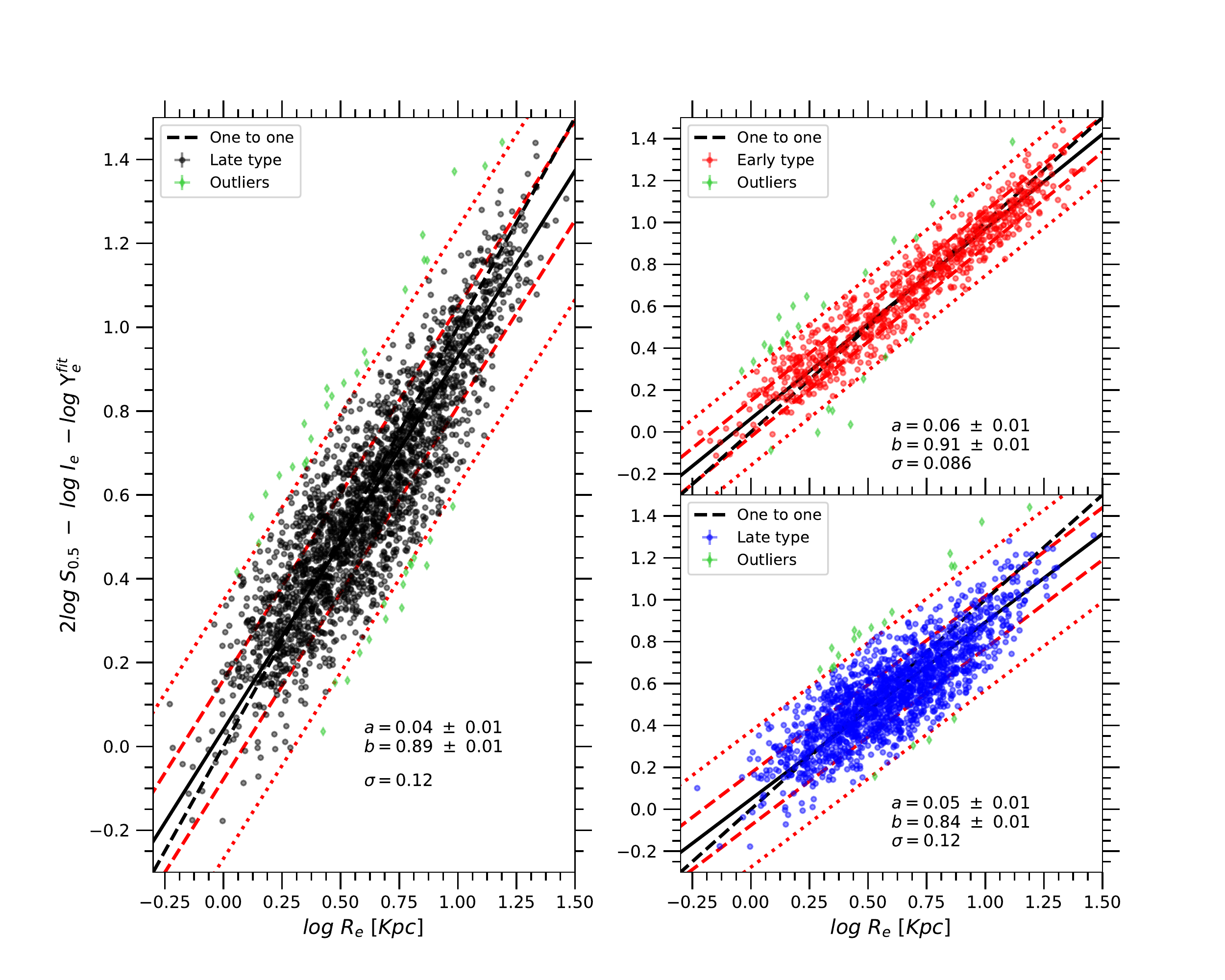}
\caption{The Fundamental Manifold for the MaNGA sample. \textit{Left:} the full sample with the dashed black line as the one-to-one relationship and the solid line the best-fit of the data. \textit{Top right)} Early type galaxies. \textit{Bottom right)} Late type galaxies. Red dashed and dotted lines marks the $1\sigma$ and $2\sigma$ bands, respectively, while green symbols as the outliers.}
\label{fig:MaNGA_FM}
\end{figure*}

We re-calibrate the $\alpha_{i}$ coefficients in Eq. \ref{eq:new} using the independent estimations of $\Upsilon_{e}^{Sch},\ S_{0.5}\ and\ I_{e}$ (defined in Section \ref{GP}) for the 300 galaxies from the CALIFA survey. For this calibration we apply a full quadratic 3D fit to the dataset. The best fit, with a reduced $\chi^{2}=0.84$, yields the following values: $\alpha_{1}=-0.60\pm0.17,\ \alpha_{2}=2.12\pm1.53,\ \alpha_{3}=-1.36\pm0.14,\ \alpha_{4}=-0.40\pm0.1,\ \alpha_{5}=0.13\pm0.04$ and $\alpha_{6}=0.20\pm0.1$.
Once the $\alpha_{i}$ coefficients have been calibrated, we apply them to explore the FM for the MaNGA sample (see Figure \ref{fig:MaNGA_FM}). We find that all galaxy types follow the same FM with a scatter of $\sim 0.1\ dex$, in great agreement with \citet{Zaritsky2008,Zaritsky2012}. Moreover, we confirm that there is a gradient in the scatter as function of the gas fraction. Thus, galaxies with low gas-fraction are located on the inner part of the FM distribution(within the $1\sigma$), while the systems with higher gas fractions are dominating the scatter. 

\section{Scaling relations for the complete MPL-7}
\label{apx:MPL9-dataset}

The increasing amount of observational data at intermediate and high redshifts have allowed to study the evolution of the TF relation, as well as of the $M_{\star}-S_{0.5}$ relation \citep[e.g.,][]{Kassin2007ApJ}. There is as of yet no convergence on the results. Some authors reported no significant evolutions \citep[e.g.,][]{Miller2011ApJ} even for the $M_{\star}-S_{0.5}$ relation \citep[e.g.,][]{Kassin2007ApJ}. Other ones reveals an evolution of the TF relation zero-points \citep[e.g.,][]{Ubler2017ApJ}. However, including all types of galaxies (merger, interacting, perturbed, face-on, edge-on, even with low signal-to-noise) i.e., without performing a cleaning of the sample, can be reflected in a variation in the slope and zero-points of scaling relations. In this Appendix we explore how the best-fitted parameters on the $M_{\star}-S_{0.5}$ relation and the Universal Fundamental Plane could be affected by making or not a detailed selection of the galaxies and spaxels within galaxies. For this analysis we have three samples:

\begin{itemize}
    \item The sample A: the full MaNGA Product Launch-7 (MPL-7) which comprise 4817 galaxies without any cleaning, i.e., it includes mergers, interacting, all environments, face and egde-on, as well as low signal-to-noise galaxies.
    \item The sample B: in this sample we apply the analysis presented in Section \ref{subsec:sample_selec} to exclude face-on, edge-on galaxies with inclinations $25^{\circ} < i > 75^{\circ}$ as well as the galaxies where the percentage of good spaxels within $R_e$ is below to 60\%, but it includes mergers and interacting galaxies. This sample comprise 2904 galaxies.
    \item The sample C: this sample is the one used along the main body of this study. Comprises the 2458 galaxies from the MPL-7 with a rigorous cleaning excluding the same that the sample B but also mergers and interacting galaxies. 
\end{itemize}

\begin{figure*}[ht!]
\centering
\includegraphics[width=1.0\textwidth, height=9cm]{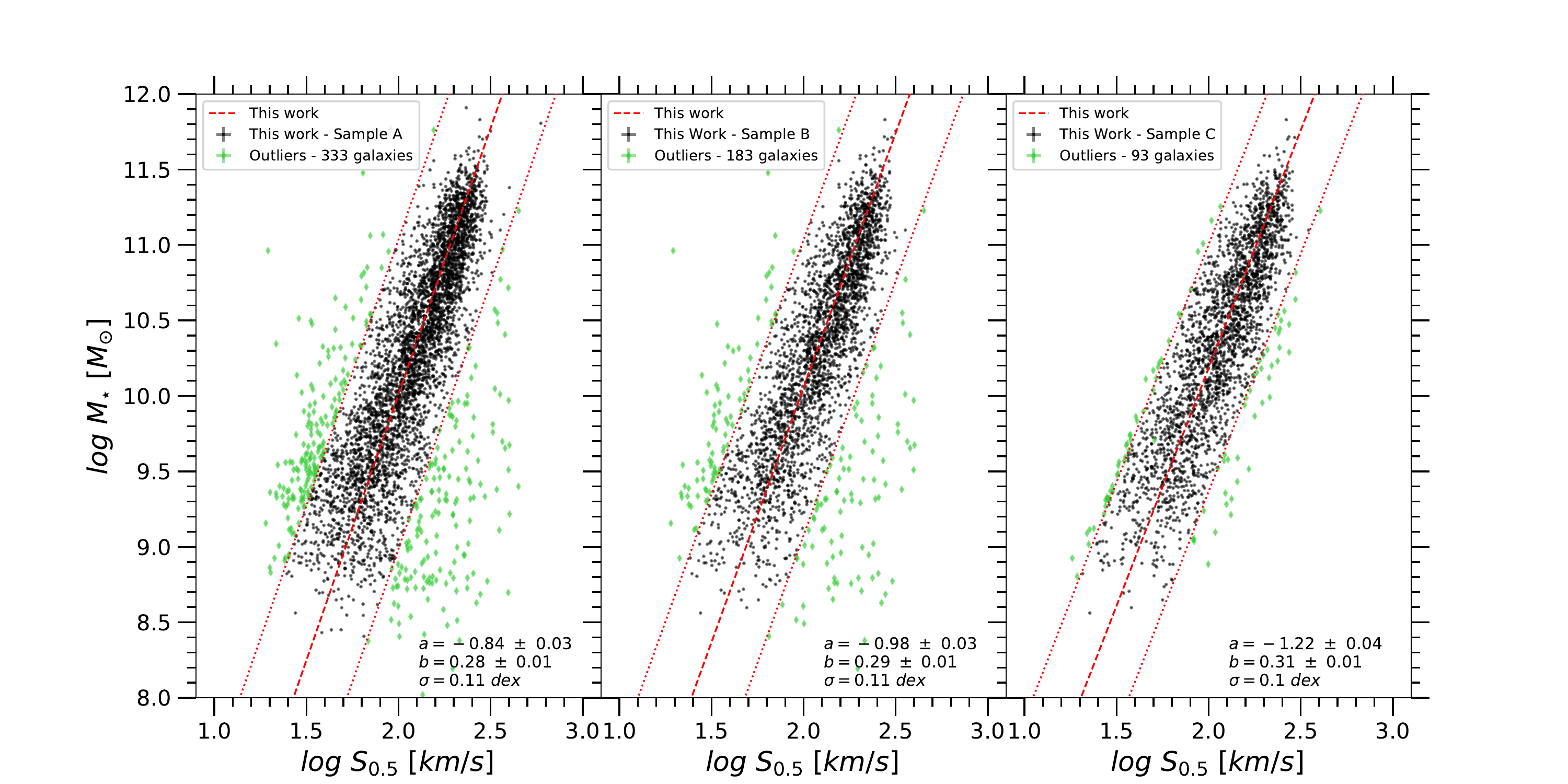}
\caption{The $M_{\star}-S_{0.5}$ relation. \textit{Left panel:} Sample A. \textit{Middle:} Sample B. \textit{Right:} Sample C. In all panels the red dashed and dotted lines marks the best-fit and the $2\sigma$ band. Green symbols are the outliers. This plot shows variations the zero-point and an increase of outliers between the rigorous depured sample an the ones poorly depured. This result could be important for high redshift studies.}
\label{fig:MaNGA_SK_MPL7}
\end{figure*}

\begin{figure}[ht!]
\centering
\includegraphics[width=1.0\textwidth, height=9cm]{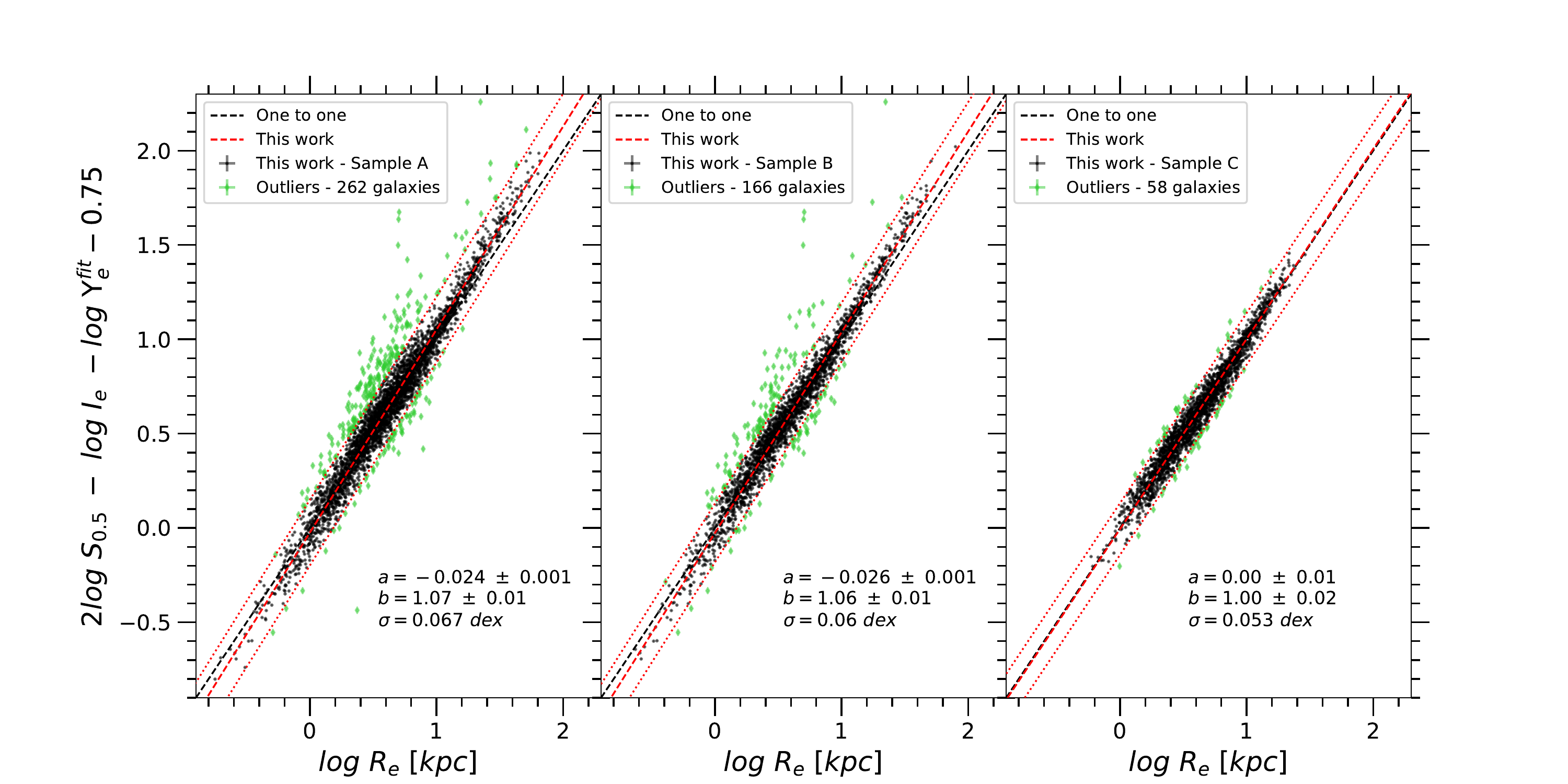}
\caption{ The Universal Fundamental Plane. \textit{Left panel:} Sample A. \textit{Middle:} Sample B. \textit{Right:} Sample C. In all panels the red dashed and dotted lines as well as green symbols are the same that Figure \ref{fig:MaNGA_SK_MPL7}.}
\label{fig:MaNGA_UFP_MPL7}
\end{figure}

In Figures \ref{fig:MaNGA_SK_MPL7} and \ref{fig:MaNGA_UFP_MPL7} we show the $M_{\star}-S_{0.5}$ relation and the Universal Fundamental Plane for the three samples mentioned above. We find slightly variations in the slope, while the variation in the zero-points are more important among the samples. The outliers increases when the sample is poorly depured without. In the $M_{\star}-S_{0.5}$ there are not variations in the scatter ($\sim 0.1\ dex$), while for the Universal Fundamental Plane the scatter increases from $0.053\ dex$ in the sample C, to $0.067\ dex$ in the sample A. This result could be important for high redshift studies.

\bibliographystyle{aasjournal}
\bibliography{referencias}

\begin{thebibliography}{}
\expandafter\ifx\csname natexlab\endcsname\relax\def\natexlab#1{#1}\fi
\providecommand{\url}[1]{\href{#1}{#1}}
\providecommand{\dodoi}[1]{doi:~\href{http://doi.org/#1}{\nolinkurl{#1}}}
\providecommand{\doeprint}[1]{\href{http://ascl.net/#1}{\nolinkurl{http://ascl.net/#1}}}
\providecommand{\doarXiv}[1]{\href{https://arxiv.org/abs/#1}{\nolinkurl{https://arxiv.org/abs/#1}}}

\bibitem[{{Akritas} \& {Bershady}(1996)}]{Akritas1996ApJ}
{Akritas}, M.~G., \& {Bershady}, M.~A. 1996, \apj, 470, 706,
  \dodoi{10.1086/177901}

\bibitem[{{Aquino-Ort{\'{\i}}z} {et~al.}(2018){Aquino-Ort{\'{\i}}z},
  {Valenzuela}, {S{\'a}nchez}, {Hern{\'a}ndez-Toledo}, {{\'A}vila-Reese}, {van
  de Ven}, {Rodr{\'{\i}}guez-Puebla}, {Zhu}, {Mancillas}, {Cano-D{\'{\i}}az},
  \& {Garc{\'{\i}}a-Benito}}]{Aquino-Ortiz2018}
{Aquino-Ort{\'{\i}}z}, E., {Valenzuela}, O., {S{\'a}nchez}, S.~F., {et~al.}
  2018, \mnras, 479, 2133, \dodoi{10.1093/mnras/sty1522}

\bibitem[{{Avila-Reese} {et~al.}(2008){Avila-Reese}, {Zavala}, {Firmani}, \&
  {Hern{\'a}ndez-Toledo}}]{Avila-Reese2008}
{Avila-Reese}, V., {Zavala}, J., {Firmani}, C., \& {Hern{\'a}ndez-Toledo},
  H.~M. 2008, \aj, 136, 1340, \dodoi{10.1088/0004-6256/136/3/1340}

\bibitem[{{Bacon} {et~al.}(2001){Bacon}, {Copin}, {Monnet}, {Miller},
  {Allington-Smith}, {Bureau}, {Carollo}, {Davies}, {Emsellem}, {Kuntschner},
  {Peletier}, {Verolme}, \& {de Zeeuw}}]{Bacon2001}
{Bacon}, R., {Copin}, Y., {Monnet}, G., {et~al.} 2001, \mnras, 326, 23,
  \dodoi{10.1046/j.1365-8711.2001.04612.x}

\bibitem[{{Bakos} \& {Trujillo}(2012)}]{Bakos2012}
{Bakos}, J., \& {Trujillo}, I. 2012, arXiv e-prints, arXiv:1204.3082.
\newblock \doarXiv{1204.3082}

\bibitem[{{Barat} {et~al.}(2019){Barat}, {D'Eugenio}, {Colless}, {Brough},
  {Catinella}, {Cortese}, {Croom}, {Medling}, {Oh}, {van de Sande}, {Sweet},
  {Yi}, {Bland-Hawthorn}, {Bryant}, {Goodwin}, {Groves}, {Lawrence}, {Owers},
  {Richards}, \& {Scott}}]{Barat2019MNRAS}
{Barat}, D., {D'Eugenio}, F., {Colless}, M., {et~al.} 2019, \mnras, 487, 2924,
  \dodoi{10.1093/mnras/stz1439}

\bibitem[{{Barrera-Ballesteros}
  {et~al.}(2015{\natexlab{a}}){Barrera-Ballesteros}, {Garc{\'\i}a-Lorenzo},
  {Falc{\'o}n-Barroso}, {van de Ven}, {Lyubenova}, {Wild}, {M{\'e}ndez-Abreu},
  {S{\'a}nchez}, {Marquez}, {Masegosa}, {Monreal-Ibero}, {Ziegler}, {del Olmo},
  {Verdes-Montenegro}, {Garc{\'\i}a-Benito}, {Husemann}, {Mast}, {Kehrig},
  {Iglesias-Paramo}, {Marino}, {Aguerri}, {Walcher}, {V{\'\i}lchez}, {Bomans},
  {Cortijo-Ferrero}, {Gonz{\'a}lez Delgado}, {Bland-Hawthorn}, {McIntosh}, \&
  {Bekerait{\.{e}}}}]{Barrera-Ballesteros2015A&A}
{Barrera-Ballesteros}, J.~K., {Garc{\'\i}a-Lorenzo}, B., {Falc{\'o}n-Barroso},
  J., {et~al.} 2015{\natexlab{a}}, \aap, 582, A21,
  \dodoi{10.1051/0004-6361/201424935}

\bibitem[{{Barrera-Ballesteros}
  {et~al.}(2015{\natexlab{b}}){Barrera-Ballesteros}, {S{\'a}nchez},
  {Garc{\'\i}a-Lorenzo}, {Falc{\'o}n-Barroso}, {Mast}, {Garc{\'\i}a-Benito},
  {Husemann}, {van de Ven}, {Iglesias-P{\'a}ramo}, {Rosales-Ortega},
  {P{\'e}rez-Torres}, {M{\'a}rquez}, {Kehrig}, {Marino}, {Vilchez}, {Galbany},
  {L{\'o}pez-S{\'a}nchez}, {Walcher}, \& {Califa
  Collaboration}}]{Barrera-Ballesteros2015A&Ab}
{Barrera-Ballesteros}, J.~K., {S{\'a}nchez}, S.~F., {Garc{\'\i}a-Lorenzo}, B.,
  {et~al.} 2015{\natexlab{b}}, \aap, 579, A45,
  \dodoi{10.1051/0004-6361/201425397}

\bibitem[{{Bekerait{\'e}} {et~al.}(2016){Bekerait{\'e}}, {Walcher},
  {Falc{\'o}n-Barroso}, {Garcia Lorenzo}, {Lyubenova}, {S{\'a}nchez},
  {Spekkens}, {van de Ven}, {Wisotzki}, {Ziegler}, {Aguerri},
  {Barrera-Ballesteros}, {Bland-Hawthorn}, {Catal{\'a}n-Torrecilla}, \&
  {Garc{\'{\i}}a-Benito}}]{Bekeraite2016}
{Bekerait{\'e}}, S., {Walcher}, C.~J., {Falc{\'o}n-Barroso}, J., {et~al.} 2016,
  \aap, 593, A114, \dodoi{10.1051/0004-6361/201527405}

\bibitem[{{Bertin} {et~al.}(2002){Bertin}, {Ciotti}, \& {Del
  Principe}}]{Bertin2002}
{Bertin}, G., {Ciotti}, L., \& {Del Principe}, M. 2002, \aap, 386, 149,
  \dodoi{10.1051/0004-6361:20020248}

\bibitem[{{Bertola} {et~al.}(1991){Bertola}, {Bettoni}, {Danziger}, {Sadler},
  {Sparke}, \& {de Zeeuw}}]{Bertola1991}
{Bertola}, F., {Bettoni}, D., {Danziger}, J., {et~al.} 1991, \apj, 373, 369,
  \dodoi{10.1086/170058}

\bibitem[{{Blanton} {et~al.}(2011){Blanton}, {Kazin}, {Muna}, {Weaver}, \&
  {Price-Whelan}}]{Blanton2011}
{Blanton}, M.~R., {Kazin}, E., {Muna}, D., {Weaver}, B.~A., \& {Price-Whelan},
  A. 2011, \aj, 142, 31, \dodoi{10.1088/0004-6256/142/1/31}

\bibitem[{{Blanton} {et~al.}(2017){Blanton}, {Bershady}, {Abolfathi},
  {Albareti}, {Allende Prieto}, {Almeida}, {Alonso-Garc{\'{\i}}a}, {Anders},
  {Anderson}, {Andrews}, \& et~al.}]{Blanton2017}
{Blanton}, M.~R., {Bershady}, M.~A., {Abolfathi}, B., {et~al.} 2017, \aj, 154,
  28, \dodoi{10.3847/1538-3881/aa7567}

\bibitem[{{Borriello} {et~al.}(2003){Borriello}, {Salucci}, \&
  {Danese}}]{Borriello2003}
{Borriello}, A., {Salucci}, P., \& {Danese}, L. 2003, \mnras, 341, 1109,
  \dodoi{10.1046/j.1365-8711.2003.06404.x}

\bibitem[{{Bundy} {et~al.}(2015){Bundy}, {Bershady}, {Law}, {Yan}, {Drory},
  {MacDonald}, {Wake}, {Cherinka}, {S{\'a}nchez-Gallego}, {Weijmans}, {Thomas},
  {Tremonti}, {Masters}, {Coccato}, {Diamond-Stanic}, {Arag{\'o}n-Salamanca},
  {Avila-Reese}, {Badenes}, {Falc{\'o}n-Barroso}, {Belfiore}, {Bizyaev},
  {Blanc}, {Bland-Hawthorn}, {Blanton}, {Brownstein}, {Byler}, {Cappellari},
  {Conroy}, {Dutton}, {Emsellem}, {Etherington}, {Frinchaboy}, {Fu}, {Gunn},
  {Harding}, {Johnston}, {Kauffmann}, {Kinemuchi}, {Klaene}, {Knapen},
  {Leauthaud}, {Li}, {Lin}, {Maiolino}, {Malanushenko}, {Malanushenko}, {Mao},
  {Maraston}, {McDermid}, {Merrifield}, {Nichol}, {Oravetz}, {Pan}, {Parejko},
  {Sanchez}, {Schlegel}, {Simmons}, {Steele}, {Steinmetz}, {Thanjavur},
  {Thompson}, {Tinker}, {van den Bosch}, {Westfall}, {Wilkinson}, {Wright},
  {Xiao}, \& {Zhang}}]{Bundy2015}
{Bundy}, K., {Bershady}, M.~A., {Law}, D.~R., {et~al.} 2015, \apj, 798, 7,
  \dodoi{10.1088/0004-637X/798/1/7}

\bibitem[{{Busarello} {et~al.}(1997){Busarello}, {Capaccioli}, {Capozziello},
  {Longo}, \& {Puddu}}]{Busarello1997}
{Busarello}, G., {Capaccioli}, M., {Capozziello}, S., {Longo}, G., \& {Puddu},
  E. 1997, \aap, 320, 415

\bibitem[{{Cappellari} \& {Emsellem}(2004)}]{CappellariPPXF2004PASP}
{Cappellari}, M., \& {Emsellem}, E. 2004, \pasp, 116, 138,
  \dodoi{10.1086/381875}

\bibitem[{{Cappellari} {et~al.}(2011){Cappellari}, {Emsellem}, {Krajnovi{\'c}},
  {McDermid}, {Scott}, {Verdoes Kleijn}, {Young}, {Alatalo}, {Bacon}, {Blitz},
  {Bois}, {Bournaud}, {Bureau}, {Davies}, {Davis}, {de Zeeuw}, {Duc},
  {Khochfar}, {Kuntschner}, {Lablanche}, {Morganti}, {Naab}, {Oosterloo},
  {Sarzi}, {Serra}, \& {Weijmans}}]{Cappellari2011MNRAS}
{Cappellari}, M., {Emsellem}, E., {Krajnovi{\'c}}, D., {et~al.} 2011, \mnras,
  413, 813, \dodoi{10.1111/j.1365-2966.2010.18174.x}

\bibitem[{{Cappellari} {et~al.}(2013){Cappellari}, {Scott}, {Alatalo}, {Blitz},
  {Bois}, {Bournaud}, {Bureau}, {Crocker}, {Davies}, {Davis}, {de Zeeuw},
  {Duc}, {Emsellem}, {Khochfar}, {Krajnovi{\'c}}, {Kuntschner}, {McDermid},
  {Morganti}, {Naab}, {Oosterloo}, {Sarzi}, {Serra}, {Weijmans}, \&
  {Young}}]{Cappellari2013}
{Cappellari}, M., {Scott}, N., {Alatalo}, K., {et~al.} 2013, \mnras, 432, 1709,
  \dodoi{10.1093/mnras/stt562}

\bibitem[{{Catinella} {et~al.}(2005){Catinella}, {Haynes}, \&
  {Giovanelli}}]{Catinella2005AJ}
{Catinella}, B., {Haynes}, M.~P., \& {Giovanelli}, R. 2005, \aj, 130, 1037,
  \dodoi{10.1086/432543}

\bibitem[{{Chabrier}(2003)}]{Chabrier2003}
{Chabrier}, G. 2003, \pasp, 115, 763, \dodoi{10.1086/376392}

\bibitem[{{Cid Fernandes} {et~al.}(2013){Cid Fernandes}, {P{\'e}rez},
  {Garc{\'{\i}}a Benito}, {Gonz{\'a}lez Delgado}, {de Amorim}, {S{\'a}nchez},
  {Husemann}, {Falc{\'o}n Barroso}, {S{\'a}nchez-Bl{\'a}zquez}, {Walcher}, \&
  {Mast}}]{Cid_Fernandes2013}
{Cid Fernandes}, R., {P{\'e}rez}, E., {Garc{\'{\i}}a Benito}, R., {et~al.}
  2013, \aap, 557, A86, \dodoi{10.1051/0004-6361/201220616}

\bibitem[{{Ciotti} {et~al.}(1996){Ciotti}, {Lanzoni}, \&
  {Renzini}}]{Ciotti1996}
{Ciotti}, L., {Lanzoni}, B., \& {Renzini}, A. 1996, \mnras, 282, 1,
  \dodoi{10.1093/mnras/282.1.1}

\bibitem[{{Cole} {et~al.}(1994){Cole}, {Aragon-Salamanca}, {Frenk}, {Navarro},
  \& {Zepf}}]{cole1994}
{Cole}, S., {Aragon-Salamanca}, A., {Frenk}, C.~S., {Navarro}, J.~F., \&
  {Zepf}, S.~E. 1994, \mnras, 271, 781, \dodoi{10.1093/mnras/271.4.781}

\bibitem[{{Cortese} {et~al.}(2014){Cortese}, {Fogarty}, {Ho}, {Bekki},
  {Bland-Hawthorn}, {Colless}, {Couch}, {Croom}, {Glazebrook}, {Mould},
  {Scott}, {Sharp}, {Tonini}, {Allen}, {Bloom}, {Bryant}, {Cluver}, {Davies},
  {Drinkwater}, {Goodwin}, {Green}, {Kewley}, {Kostantopoulos}, {Lawrence},
  {Mahajan}, {Medling}, {Owers}, {Richards}, {Sweet}, \& {Wong}}]{Cortese2014}
{Cortese}, L., {Fogarty}, L.~M.~R., {Ho}, I.-T., {et~al.} 2014, \apjl, 795,
  L37, \dodoi{10.1088/2041-8205/795/2/L37}

\bibitem[{{Courteau} {et~al.}(2007){Courteau}, {Dutton}, {van den Bosch},
  {MacArthur}, {Dekel}, {McIntosh}, \& {Dale}}]{Courteau2007}
{Courteau}, S., {Dutton}, A.~A., {van den Bosch}, F.~C., {et~al.} 2007, \apj,
  671, 203, \dodoi{10.1086/522193}

\bibitem[{{Courteau} \& {Rix}(1999)}]{Courteau&Rix1999}
{Courteau}, S., \& {Rix}, H.-W. 1999, \apj, 513, 561, \dodoi{10.1086/306872}

\bibitem[{{Courteau} {et~al.}(2014){Courteau}, {Cappellari}, {de Jong},
  {Dutton}, {Emsellem}, {Hoekstra}, {Koopmans}, {Mamon}, {Maraston}, {Treu}, \&
  {Widrow}}]{Courteau2014}
{Courteau}, S., {Cappellari}, M., {de Jong}, R.~S., {et~al.} 2014, Reviews of
  Modern Physics, 86, 47, \dodoi{10.1103/RevModPhys.86.47}

\bibitem[{{Croom} {et~al.}(2012){Croom}, {Lawrence}, {Bland-Hawthorn},
  {Bryant}, {Fogarty}, {Richards}, {Goodwin}, {Farrell}, {Miziarski}, {Heald},
  {Jones}, {Lee}, {Colless}, {Brough}, {Hopkins}, {Bauer}, {Birchall}, {Ellis},
  {Horton}, {Leon-Saval}, {Lewis}, {L{\'o}pez-S{\'a}nchez}, {Min}, {Trinh}, \&
  {Trowland}}]{Croom2012MNRAS}
{Croom}, S.~M., {Lawrence}, J.~S., {Bland-Hawthorn}, J., {et~al.} 2012, \mnras,
  421, 872, \dodoi{10.1111/j.1365-2966.2011.20365.x}

\bibitem[{{De Rossi} {et~al.}(2012){De Rossi}, {Tissera}, \&
  {Pedrosa}}]{DeRossi2012}
{De Rossi}, M.~E., {Tissera}, P.~B., \& {Pedrosa}, S.~E. 2012, \aap, 546, A52,
  \dodoi{10.1051/0004-6361/201118409}

\bibitem[{{DESI Collaboration} {et~al.}(2016){DESI Collaboration}, {Aghamousa},
  {Aguilar}, {Ahlen}, {Alam}, {Allen}, {Allende Prieto}, {Annis}, {Bailey},
  {Balland}, {Ballester}, {Baltay}, {Beaufore}, {Bebek}, {Beers}, {Bell},
  {Bernal}, {Besuner}, {Beutler}, {Blake}, {Bleuler}, {Blomqvist}, {Blum},
  {Bolton}, {Briceno}, {Brooks}, {Brownstein}, {Buckley-Geer}, {Burden},
  {Burtin}, {Busca}, {Cahn}, {Cai}, {Cardiel-Sas}, {Carlberg}, {Carton},
  {Casas}, {Castand er}, {Cervantes-Cota}, {Claybaugh}, {Close}, {Coker},
  {Cole}, {Comparat}, {Cooper}, {Cousinou}, {Crocce}, {Cuby}, {Cunningham},
  {Davis}, {Dawson}, {de la Macorra}, {De Vicente}, {Delubac}, {Derwent},
  {Dey}, {Dhungana}, {Ding}, {Doel}, {Duan}, {Ealet}, {Edelstein},
  {Eftekharzadeh}, {Eisenstein}, {Elliott}, {Escoffier}, {Evatt}, {Fagrelius},
  {Fan}, {Fanning}, {Farahi}, {Farihi}, {Favole}, {Feng}, {Fernandez},
  {Findlay}, {Finkbeiner}, {Fitzpatrick}, {Flaugher}, {Flender}, {Font-Ribera},
  {Forero-Romero}, {Fosalba}, {Frenk}, {Fumagalli}, {Gaensicke}, {Gallo},
  {Garcia-Bellido}, {Gaztanaga}, {Pietro Gentile Fusillo}, {Gerard},
  {Gershkovich}, {Giannantonio}, {Gillet}, {Gonzalez-de-Rivera},
  {Gonzalez-Perez}, {Gott}, {Graur}, {Gutierrez}, {Guy}, {Habib}, {Heetderks},
  {Heetderks}, {Heitmann}, {Hellwing}, {Herrera}, {Ho}, {Holland}, {Honscheid},
  {Huff}, {Hutchinson}, {Huterer}, {Hwang}, {Illa Laguna}, {Ishikawa},
  {Jacobs}, {Jeffrey}, {Jelinsky}, {Jennings}, {Jiang}, {Jimenez}, {Johnson},
  {Joyce}, {Jullo}, {Juneau}, {Kama}, {Karcher}, {Karkar}, {Kehoe}, {Kennamer},
  {Kent}, {Kilbinger}, {Kim}, {Kirkby}, {Kisner}, {Kitanidis}, {Kneib},
  {Koposov}, {Kovacs}, {Koyama}, {Kremin}, {Kron}, {Kronig}, {Kueter-Young},
  {Lacey}, {Lafever}, {Lahav}, {Lambert}, {Lampton}, {Land riau}, {Lang},
  {Lauer}, {Le Goff}, {Le Guillou}, {Le Van Suu}, {Lee}, {Lee}, {Leitner},
  {Lesser}, {Levi}, {L'Huillier}, {Li}, {Liang}, {Lin}, {Linder}, {Loebman},
  {Luki{\'c}}, {Ma}, {MacCrann}, {Magneville}, {Makarem}, {Manera}, {Manser},
  {Marshall}, {Martini}, {Massey}, {Matheson}, {McCauley}, {McDonald},
  {McGreer}, {Meisner}, {Metcalfe}, {Miller}, {Miquel}, {Moustakas}, {Myers},
  {Naik}, {Newman}, {Nichol}, {Nicola}, {Nicolati da Costa}, {Nie}, {Niz},
  {Norberg}, {Nord}, {Norman}, {Nugent}, {O'Brien}, {Oh}, {Olsen}, {Padilla},
  {Padmanabhan}, {Padmanabhan}, {Palanque-Delabrouille}, {Palmese},
  {Pappalardo}, {P{\^a}ris}, {Park}, {Patej}, {Peacock}, {Peiris}, {Peng},
  {Percival}, {Perruchot}, {Pieri}, {Pogge}, {Pollack}, {Poppett}, {Prada},
  {Prakash}, {Probst}, {Rabinowitz}, {Raichoor}, {Ree}, {Refregier}, {Regal},
  {Reid}, {Reil}, {Rezaie}, {Rockosi}, {Roe}, {Ronayette}, {Roodman}, {Ross},
  {Ross}, {Rossi}, {Rozo}, {Ruhlmann-Kleider}, {Rykoff}, {Sabiu}, {Samushia},
  {Sanchez}, {Sanchez}, {Schlegel}, {Schneider}, {Schubnell}, {Secroun},
  {Seljak}, {Seo}, {Serrano}, {Shafieloo}, {Shan}, {Sharples}, {Sholl},
  {Shourt}, {Silber}, {Silva}, {Sirk}, {Slosar}, {Smith}, {Smoot}, {Som},
  {Song}, {Sprayberry}, {Staten}, {Stefanik}, {Tarle}, {Sien Tie}, {Tinker},
  {Tojeiro}, {Valdes}, {Valenzuela}, {Valluri}, {Vargas-Magana}, {Verde},
  {Walker}, {Wang}, {Wang}, {Weaver}, {Weaverdyck}, {Wechsler}, {Weinberg},
  {White}, {Yang}, {Yeche}, {Zhang}, {Zhao}, {Zheng}, {Zhou}, {Zhou}, {Zhu},
  {Zou}, \& {Zu}}]{DESI2016}
{DESI Collaboration}, {Aghamousa}, A., {Aguilar}, J., {et~al.} 2016, arXiv
  e-prints, arXiv:1611.00036.
\newblock \doarXiv{1611.00036}

\bibitem[{{Djorgovski} \& {Davis}(1987)}]{Djorgovski&Davis1987}
{Djorgovski}, S., \& {Davis}, M. 1987, \apj, 313, 59, \dodoi{10.1086/164948}

\bibitem[{{Dressler} {et~al.}(1987){Dressler}, {Lynden-Bell}, {Burstein},
  {Davies}, {Faber}, {Terlevich}, \& {Wegner}}]{Dressler1987}
{Dressler}, A., {Lynden-Bell}, D., {Burstein}, D., {et~al.} 1987, \apj, 313,
  42, \dodoi{10.1086/164947}

\bibitem[{{Drory} {et~al.}(2015){Drory}, {MacDonald}, {Bershady}, {Bundy},
  {Gunn}, {Law}, {Smith}, {Stoll}, {Tremonti}, {Wake}, {Yan}, {Weijmans},
  {Byler}, {Cherinka}, {Cope}, {Eigenbrot}, {Harding}, {Holder}, {Huehnerhoff},
  {Jaehnig}, {Jansen}, {Klaene}, {Paat}, {Percival}, \& {Sayres}}]{Drory2015}
{Drory}, N., {MacDonald}, N., {Bershady}, M.~A., {et~al.} 2015, \aj, 149, 77,
  \dodoi{10.1088/0004-6256/149/2/77}

\bibitem[{{Dutton} {et~al.}(2013){Dutton}, {Macci{\`o}}, {Mendel}, \&
  {Simard}}]{Dutton2013}
{Dutton}, A.~A., {Macci{\`o}}, A.~V., {Mendel}, J.~T., \& {Simard}, L. 2013,
  \mnras, 432, 2496, \dodoi{10.1093/mnras/stt608}

\bibitem[{{Emsellem} {et~al.}(2007){Emsellem}, {Cappellari}, {Krajnovi{\'c}},
  {van de Ven}, {Bacon}, {Bureau}, {Davies}, {de Zeeuw}, {Falc{\'o}n-Barroso},
  {Kuntschner}, {McDermid}, {Peletier}, \& {Sarzi}}]{Emsellem2007MNRAS}
{Emsellem}, E., {Cappellari}, M., {Krajnovi{\'c}}, D., {et~al.} 2007, \mnras,
  379, 401, \dodoi{10.1111/j.1365-2966.2007.11752.x}

\bibitem[{{Espinosa-Ponce} {et~al.}(2020){Espinosa-Ponce}, {S{\'a}nchez},
  {Morisset}, {Barrera-Ballesteros}, {Galbany}, {Garc{\'\i}a-Benito},
  {Lacerda}, \& {Mast}}]{Espinosa-Ponce2020}
{Espinosa-Ponce}, C., {S{\'a}nchez}, S.~F., {Morisset}, C., {et~al.} 2020,
  \mnras, 494, 1622, \dodoi{10.1093/mnras/staa782}

\bibitem[{{Faber} {et~al.}(1987){Faber}, {Dressler}, {Davies}, {Burstein},
  {Lynden Bell}, {Terlevich}, \& {Wegner}}]{Faber1987}
{Faber}, S.~M., {Dressler}, A., {Davies}, R.~L., {et~al.} 1987, in Nearly
  Normal Galaxies. From the Planck Time to the Present, ed. S.~M. {Faber}, 175

\bibitem[{{Faber} \& {Jackson}(1976)}]{Faber-Jackson1976}
{Faber}, S.~M., \& {Jackson}, R.~E. 1976, \apj, 204, 668,
  \dodoi{10.1086/154215}

\bibitem[{{Falc{\'o}n-Barroso} {et~al.}(2011){Falc{\'o}n-Barroso}, {van de
  Ven}, {Peletier}, {Bureau}, {Jeong}, {Bacon}, {Cappellari}, {Davies}, {de
  Zeeuw}, {Emsellem}, {Krajnovi{\'c}}, {Kuntschner}, {McDermid}, {Sarzi},
  {Shapiro}, {van den Bosch}, {van der Wolk}, {Weijmans}, \&
  {Yi}}]{Falcon-Barroso2011}
{Falc{\'o}n-Barroso}, J., {van de Ven}, G., {Peletier}, R.~F., {et~al.} 2011,
  \mnras, 417, 1787, \dodoi{10.1111/j.1365-2966.2011.19372.x}

\bibitem[{{Falc{\'o}n-Barroso} {et~al.}(2017){Falc{\'o}n-Barroso}, {Lyubenova},
  {van de Ven}, {Mendez-Abreu}, {Aguerri}, {Garc{\'\i}a-Lorenzo},
  {Bekerait{\'e}}, {S{\'a}nchez}, {Husemann}, {Garc{\'\i}a-Benito}, {Mast},
  {Walcher}, {Zibetti}, {Barrera-Ballesteros}, {Galbany},
  {S{\'a}nchez-Bl{\'a}zquez}, {Singh}, {van den Bosch}, {Wild}, {Zhu}, {Bland
  -Hawthorn}, {Cid Fernandes}, {de Lorenzo-C{\'a}ceres}, {Gallazzi},
  {Gonz{\'a}lez Delgado}, {Marino}, {M{\'a}rquez}, {P{\'e}rez}, {P{\'e}rez},
  {Roth}, {Rosales-Ortega}, {Ruiz-Lara}, {Wisotzki}, {Ziegler}, \& {Califa
  Collaboration}}]{Falcon-Barroso2017}
{Falc{\'o}n-Barroso}, J., {Lyubenova}, M., {van de Ven}, G., {et~al.} 2017,
  \aap, 597, A48, \dodoi{10.1051/0004-6361/201628625}

\bibitem[{{Falc{\'o}n-Barroso} {et~al.}(2019){Falc{\'o}n-Barroso}, {van de
  Ven}, {Lyubenova}, {Mendez-Abreu}, {Aguerri}, {Garc{\'\i}a-Lorenzo},
  {Bekerait{\'e}}, {S{\'a}nchez}, {Husemann}, {Garc{\'\i}a-Benito},
  {Gonz{\'a}lez Delgado}, {Mast}, {Walcher}, {Zibetti}, {Zhu},
  {Barrera-Ballesteros}, {Galbany}, {S{\'a}nchez-Bl{\'a}zquez}, {Singh}, {van
  den Bosch}, {Wild}, {Bland-Hawthorn}, {Cid Fernandes}, {de
  Lorenzo-C{\'a}ceres}, {Gallazzi}, {Marino}, {M{\'a}rquez}, {Peletier},
  {P{\'e}rez}, {P{\'e}rez}, {Roth}, {Rosales-Ortega}, {Ruiz-Lara}, {Wisotzki},
  \& {Ziegler}}]{Falcon-Barroso2019A&A}
{Falc{\'o}n-Barroso}, J., {van de Ven}, G., {Lyubenova}, M., {et~al.} 2019,
  \aap, 632, A59, \dodoi{10.1051/0004-6361/201936413}

\bibitem[{{Ferrero} {et~al.}(2017){Ferrero}, {Navarro}, {Abadi}, {Sales},
  {Bower}, {Crain}, {Frenk}, {Schaller}, {Schaye}, \& {Theuns}}]{Ferrero2017}
{Ferrero}, I., {Navarro}, J.~F., {Abadi}, M.~G., {et~al.} 2017, \mnras, 464,
  4736, \dodoi{10.1093/mnras/stw2691}

\bibitem[{{Firmani} \& {Avila-Reese}(2000)}]{Firmani2000}
{Firmani}, C., \& {Avila-Reese}, V. 2000, \mnras, 315, 457,
  \dodoi{10.1046/j.1365-8711.2000.03338.x}

\bibitem[{{Fischer} {et~al.}(2019){Fischer}, {Dom{\'\i}nguez S{\'a}nchez}, \&
  {Bernardi}}]{Fischer2019MNRAS}
{Fischer}, J.~L., {Dom{\'\i}nguez S{\'a}nchez}, H., \& {Bernardi}, M. 2019,
  \mnras, 483, 2057, \dodoi{10.1093/mnras/sty3135}

\bibitem[{{Forbes} {et~al.}(1998){Forbes}, {Ponman}, \& {Brown}}]{Forbes1998}
{Forbes}, D.~A., {Ponman}, T.~J., \& {Brown}, R.~J.~N. 1998, \apjl, 508, L43,
  \dodoi{10.1086/311715}

\bibitem[{{Freeman}(1970)}]{Freeman1970}
{Freeman}, K.~C. 1970, \apj, 160, 811, \dodoi{10.1086/150474}

\bibitem[{{Gallazzi} {et~al.}(2006){Gallazzi}, {Charlot}, {Brinchmann}, \&
  {White}}]{Gallazzi2006}
{Gallazzi}, A., {Charlot}, S., {Brinchmann}, J., \& {White}, S.~D.~M. 2006,
  \mnras, 370, 1106, \dodoi{10.1111/j.1365-2966.2006.10548.x}

\bibitem[{{Garc{\'\i}a-Benito} {et~al.}(2019){Garc{\'\i}a-Benito},
  {Gonz{\'a}lez Delgado}, {P{\'e}rez}, {Cid Fernandes}, {S{\'a}nchez}, \& {de
  Amorim}}]{Garcia-Benito2019}
{Garc{\'\i}a-Benito}, R., {Gonz{\'a}lez Delgado}, R.~M., {P{\'e}rez}, E.,
  {et~al.} 2019, \aap, 621, A120, \dodoi{10.1051/0004-6361/201833993}

\bibitem[{{Garc{\'\i}a-Benito} {et~al.}(2015){Garc{\'\i}a-Benito}, {Zibetti},
  {S{\'a}nchez}, {Husemann}, {de Amorim}, {Castillo-Morales}, {Cid Fernandes},
  {Ellis}, {Falc{\'o}n-Barroso}, {Galbany}, {Gil de Paz}, {Gonz{\'a}lez
  Delgado}, {Lacerda}, {L{\'o}pez-Fernandez}, {de Lorenzo-C{\'a}ceres},
  {Lyubenova}, {Marino}, {Mast}, {Mendoza}, {P{\'e}rez}, {Vale Asari},
  {Aguerri}, {Ascasibar}, {Bekerait{\.{e}}}, {Bland-Hawthorn},
  {Barrera-Ballesteros}, {Bomans}, {Cano-D{\'\i}az}, {Catal{\'a}n-Torrecilla},
  {Cortijo}, {Delgado-Inglada}, {Demleitner}, {Dettmar}, {D{\'\i}az},
  {Florido}, {Gallazzi}, {Garc{\'\i}a-Lorenzo}, {Gomes}, {Holmes},
  {Iglesias-P{\'a}ramo}, {Jahnke}, {Kalinova}, {Kehrig}, {Kennicutt},
  {L{\'o}pez-S{\'a}nchez}, {M{\'a}rquez}, {Masegosa}, {Meidt}, {Mendez-Abreu},
  {Moll{\'a}}, {Monreal-Ibero}, {Morisset}, {del Olmo}, {Papaderos},
  {P{\'e}rez}, {Quirrenbach}, {Rosales-Ortega}, {Roth}, {Ruiz-Lara},
  {S{\'a}nchez-Bl{\'a}zquez}, {S{\'a}nchez-Menguiano}, {Singh}, {Spekkens},
  {Stanishev}, {Torres-Papaqui}, {van de Ven}, {Vilchez}, {Walcher}, {Wild},
  {Wisotzki}, {Ziegler}, {Alves}, {Barrado}, {Quintana}, \&
  {Aceituno}}]{GarciaBenito2015A&A}
{Garc{\'\i}a-Benito}, R., {Zibetti}, S., {S{\'a}nchez}, S.~F., {et~al.} 2015,
  \aap, 576, A135, \dodoi{10.1051/0004-6361/201425080}

\bibitem[{{Gilhuly} {et~al.}(2019){Gilhuly}, {Courteau}, \&
  {S{\'a}nchez}}]{Gilhuly2019}
{Gilhuly}, C., {Courteau}, S., \& {S{\'a}nchez}, S.~F. 2019, \mnras, 482, 1427,
  \dodoi{10.1093/mnras/sty2792}

\bibitem[{{Giovanelli} {et~al.}(1997){Giovanelli}, {Haynes}, {Herter}, {Vogt},
  {da Costa}, {Freudling}, {Salzer}, \& {Wegner}}]{Giovanelli1997}
{Giovanelli}, R., {Haynes}, M.~P., {Herter}, T., {et~al.} 1997, \aj, 113, 53,
  \dodoi{10.1086/118234}

\bibitem[{{Graham} \& {Colless}(1997)}]{Graham&Colless1997}
{Graham}, A., \& {Colless}, M. 1997, \mnras, 287, 221,
  \dodoi{10.1093/mnras/287.1.221}

\bibitem[{{Graham} {et~al.}(2018){Graham}, {Cappellari}, {Li}, {Mao},
  {Bershady}, {Bizyaev}, {Brinkmann}, {Brownstein}, {Bundy}, {Drory}, {Law},
  {Pan}, {Thomas}, {Wake}, {Weijmans}, {Westfall}, \& {Yan}}]{Graham2018MNRAS}
{Graham}, M.~T., {Cappellari}, M., {Li}, H., {et~al.} 2018, \mnras, 477, 4711,
  \dodoi{10.1093/mnras/sty504}

\bibitem[{{Gunn} {et~al.}(2006){Gunn}, {Siegmund}, {Mannery}, {Owen}, {Hull},
  {Leger}, {Carey}, {Knapp}, {York}, {Boroski}, {Kent}, {Lupton}, {Rockosi},
  {Evans}, {Waddell}, {Anderson}, {Annis}, {Barentine}, {Bartoszek}, {Bastian},
  {Bracker}, {Brewington}, {Briegel}, {Brinkmann}, {Brown}, {Carr},
  {Czarapata}, {Drennan}, {Dombeck}, {Federwitz}, {Gillespie}, {Gonzales},
  {Hansen}, {Harvanek}, {Hayes}, {Jordan}, {Kinney}, {Klaene}, {Kleinman},
  {Kron}, {Kresinski}, {Lee}, {Limmongkol}, {Lindenmeyer}, {Long}, {Loomis},
  {McGehee}, {Mantsch}, {Neilsen}, {Neswold}, {Newman}, {Nitta}, {Peoples},
  {Pier}, {Prieto}, {Prosapio}, {Rivetta}, {Schneider}, {Snedden}, \&
  {Wang}}]{Gunn2006}
{Gunn}, J.~E., {Siegmund}, W.~A., {Mannery}, E.~J., {et~al.} 2006, \aj, 131,
  2332, \dodoi{10.1086/500975}

\bibitem[{{Hall} {et~al.}(2012){Hall}, {Courteau}, {Dutton}, {McDonald}, \&
  {Zhu}}]{Hall2012}
{Hall}, M., {Courteau}, S., {Dutton}, A.~A., {McDonald}, M., \& {Zhu}, Y. 2012,
  \mnras, 425, 2741, \dodoi{10.1111/j.1365-2966.2012.21290.x}

\bibitem[{{Holmes} {et~al.}(2015){Holmes}, {Spekkens}, {S{\'a}nchez},
  {Walcher}, {Garc{\'{\i}}a-Benito}, {Mast}, {Cortijo-Ferrero}, {Kalinova},
  {Marino}, {Mendez-Abreu}, \& {Barrera-Ballesteros}}]{Holmes2015}
{Holmes}, L., {Spekkens}, K., {S{\'a}nchez}, S.~F., {et~al.} 2015, \mnras, 451,
  4397, \dodoi{10.1093/mnras/stv1254}

\bibitem[{{Husemann} {et~al.}(2013){Husemann}, {Jahnke}, {S{\'a}nchez},
  {Barrado}, {Bekerait{\.{e}}}, {Bomans}, {Castillo-Morales},
  {Catal{\'a}n-Torrecilla}, {Cid Fernand es}, {Falc{\'o}n-Barroso},
  {Garc{\'\i}a-Benito}, {Gonz{\'a}lez Delgado}, {Iglesias-P{\'a}ramo},
  {Johnson}, {Kupko}, {L{\'o}pez-Fernandez}, {Lyubenova}, {Marino}, {Mast},
  {Miskolczi}, {Monreal-Ibero}, {Gil de Paz}, {P{\'e}rez}, {P{\'e}rez},
  {Rosales-Ortega}, {Ruiz-Lara}, {Schilling}, {van de Ven}, {Walcher}, {Alves},
  {de Amorim}, {Backsmann}, {Barrera-Ballesteros}, {Bland-Hawthorn}, {Cortijo},
  {Dettmar}, {Demleitner}, {D{\'\i}az}, {Enke}, {Florido}, {Flores}, {Galbany},
  {Gallazzi}, {Garc{\'\i}a-Lorenzo}, {Gomes}, {Gruel}, {Haines}, {Holmes},
  {Jungwiert}, {Kalinova}, {Kehrig}, {Kennicutt}, {Klar}, {Lehnert},
  {L{\'o}pez-S{\'a}nchez}, {de Lorenzo-C{\'a}ceres}, {M{\'a}rmol-Queralt{\'o}},
  {M{\'a}rquez}, {Mendez-Abreu}, {Moll{\'a}}, {del Olmo}, {Meidt}, {Papaderos},
  {Puschnig}, {Quirrenbach}, {Roth}, {S{\'a}nchez-Bl{\'a}zquez}, {Spekkens},
  {Singh}, {Stanishev}, {Trager}, {Vilchez}, {Wild}, {Wisotzki}, {Zibetti}, \&
  {Ziegler}}]{Husemann2013A&A}
{Husemann}, B., {Jahnke}, K., {S{\'a}nchez}, S.~F., {et~al.} 2013, \aap, 549,
  A87, \dodoi{10.1051/0004-6361/201220582}

\bibitem[{{Jin} {et~al.}(2020){Jin}, {Zhu}, {Long}, {Mao}, {Wang}, \& {van de
  Ven}}]{Yunpeng2020}
{Jin}, Y., {Zhu}, L., {Long}, R.~J., {et~al.} 2020, \mnras, 491, 1690,
  \dodoi{10.1093/mnras/stz3072}

\bibitem[{{Kalinova} {et~al.}(2017){Kalinova}, {Colombo}, {Rosolowsky},
  {Kannan}, {Galbany}, {Garc{\'\i}a-Benito}, {Gonz{\'a}lez Delgado},
  {S{\'a}nchez}, {Ruiz-Lara}, {M{\'e}ndez-Abreu}, {Catal{\'a}n-Torrecilla},
  {S{\'a}nchez-Menguiano}, {de Lorenzo-C{\'a}ceres}, {Costantin}, {Florido},
  {Kodaira}, {Marino}, {L{\"a}sker}, \& {Bland -Hawthorn}}]{Kalinova2017MNRAS}
{Kalinova}, V., {Colombo}, D., {Rosolowsky}, E., {et~al.} 2017, \mnras, 469,
  2539, \dodoi{10.1093/mnras/stx901}

\bibitem[{{Kassin} {et~al.}(2007){Kassin}, {Weiner}, {Faber}, {Koo}, {Lotz},
  {Diemand}, {Harker}, {Bundy}, {Metevier}, {Phillips}, {Cooper}, {Croton},
  {Konidaris}, {Noeske}, \& {Willmer}}]{Kassin2007ApJ}
{Kassin}, S.~A., {Weiner}, B.~J., {Faber}, S.~M., {et~al.} 2007, \apjl, 660,
  L35, \dodoi{10.1086/517932}

\bibitem[{{Kelz} {et~al.}(2006){Kelz}, {Verheijen}, {Roth}, {Bauer}, {Becker},
  {Paschke}, {Popow}, {S{\'a}nchez}, \& {Laux}}]{Kelz2006PASP}
{Kelz}, A., {Verheijen}, M. A.~W., {Roth}, M.~M., {et~al.} 2006, \pasp, 118,
  129, \dodoi{10.1086/497455}

\bibitem[{{Lacerda} {et~al.}(2020){Lacerda}, {S{\'a}nchez}, {Cid Fernandes},
  {L{\'o}pez-Cob{\'a}}, {Espinosa-Ponce}, \& {Galbany}}]{Lacerda2020MNRAS}
{Lacerda}, E. A.~D., {S{\'a}nchez}, S.~F., {Cid Fernandes}, R., {et~al.} 2020,
  \mnras, 492, 3073, \dodoi{10.1093/mnras/staa008}

\bibitem[{{Law} {et~al.}(2015){Law}, {Yan}, {Bershady}, {Bundy}, {Cherinka},
  {Drory}, {MacDonald}, {S{\'a}nchez-Gallego}, {Wake}, {Weijmans}, {Blanton},
  {Klaene}, {Moran}, {Sanchez}, \& {Zhang}}]{Law2015}
{Law}, D.~R., {Yan}, R., {Bershady}, M.~A., {et~al.} 2015, \aj, 150, 19,
  \dodoi{10.1088/0004-6256/150/1/19}

\bibitem[{{Law} {et~al.}(2016){Law}, {Cherinka}, {Yan}, {Andrews}, {Bershady},
  {Bizyaev}, {Blanc}, {Blanton}, {Bolton}, {Brownstein}, {Bundy}, {Chen},
  {Drory}, {D'Souza}, {Fu}, {Jones}, {Kauffmann}, {MacDonald}, {Masters},
  {Newman}, {Parejko}, {S{\'a}nchez-Gallego}, {S{\'a}nchez}, {Schlegel},
  {Thomas}, {Wake}, {Weijmans}, {Westfall}, \& {Zhang}}]{Law2016}
{Law}, D.~R., {Cherinka}, B., {Yan}, R., {et~al.} 2016, \aj, 152, 83,
  \dodoi{10.3847/0004-6256/152/4/83}

\bibitem[{{Li} {et~al.}(2018){Li}, {Mao}, {Cappellari}, {Ge}, {Long}, {Li},
  {Mo}, {Li}, {Zheng}, {Bundy}, {Thomas}, {Brownstein}, {Roman Lopes}, {Law},
  \& {Drory}}]{Li2018MNRAS}
{Li}, H., {Mao}, S., {Cappellari}, M., {et~al.} 2018, \mnras, 476, 1765,
  \dodoi{10.1093/mnras/sty334}

\bibitem[{{Mart{\'\i}n-Navarro} {et~al.}(2015){Mart{\'\i}n-Navarro},
  {Vazdekis}, {La Barbera}, {Falc{\'o}n-Barroso}, {Lyubenova}, {van de Ven},
  {Ferreras}, {S{\'a}nchez}, {Trager}, {Garc{\'\i}a-Benito}, {Mast}, {Mendoza},
  {S{\'a}nchez-Bl{\'a}zquez}, {Gonz{\'a}lez Delgado}, {Walcher}, \& {CALIFA
  Team}}]{Martin-Navarro2015}
{Mart{\'\i}n-Navarro}, I., {Vazdekis}, A., {La Barbera}, F., {et~al.} 2015,
  \apjl, 806, L31, \dodoi{10.1088/2041-8205/806/2/L31}

\bibitem[{{McGaugh} {et~al.}(2000){McGaugh}, {Schombert}, {Bothun}, \& {de
  Blok}}]{McGaugh2000}
{McGaugh}, S.~S., {Schombert}, J.~M., {Bothun}, G.~D., \& {de Blok}, W.~J.~G.
  2000, \apjl, 533, L99, \dodoi{10.1086/312628}

\bibitem[{{Meyer} {et~al.}(2008){Meyer}, {Zwaan}, {Webster}, {Schneider}, \&
  {Staveley-Smith}}]{Mayer2008}
{Meyer}, M.~J., {Zwaan}, M.~A., {Webster}, R.~L., {Schneider}, S., \&
  {Staveley-Smith}, L. 2008, \mnras, 391, 1712,
  \dodoi{10.1111/j.1365-2966.2008.13424.x}

\bibitem[{{Miller} {et~al.}(2011){Miller}, {Bundy}, {Sullivan}, {Ellis}, \&
  {Treu}}]{Miller2011ApJ}
{Miller}, S.~H., {Bundy}, K., {Sullivan}, M., {Ellis}, R.~S., \& {Treu}, T.
  2011, \apj, 741, 115, \dodoi{10.1088/0004-637X/741/2/115}

\bibitem[{{Mo} {et~al.}(1998){Mo}, {Mao}, \& {White}}]{Mo&Mao&White1998}
{Mo}, H.~J., {Mao}, S., \& {White}, S.~D.~M. 1998, \mnras, 295, 319,
  \dodoi{10.1046/j.1365-8711.1998.01227.x}

\bibitem[{{Padmanabhan} {et~al.}(2004){Padmanabhan}, {Seljak}, {Strauss},
  {Blanton}, {Kauffmann}, {Schlegel}, {Tremonti}, {Bahcall}, {Bernardi},
  {Brinkmann}, {Fukugita}, \& {Ivezi{\'c}}}]{Padmanabhan2004}
{Padmanabhan}, N., {Seljak}, U., {Strauss}, M.~A., {et~al.} 2004, \na, 9, 329,
  \dodoi{10.1016/j.newast.2003.12.004}

\bibitem[{{Papastergis} {et~al.}(2011){Papastergis}, {Martin}, {Giovanelli}, \&
  {Haynes}}]{Papastergis2011}
{Papastergis}, E., {Martin}, A.~M., {Giovanelli}, R., \& {Haynes}, M.~P. 2011,
  \apj, 739, 38, \dodoi{10.1088/0004-637X/739/1/38}

\bibitem[{{Pizagno} {et~al.}(2007){Pizagno}, {Prada}, {Weinberg}, {Rix},
  {Pogge}, {Grebel}, {Harbeck}, {Blanton}, {Brinkmann}, \&
  {Gunn}}]{Pizagno2007}
{Pizagno}, J., {Prada}, F., {Weinberg}, D.~H., {et~al.} 2007, \aj, 134, 945,
  \dodoi{10.1086/519522}

\bibitem[{{Ponomareva} {et~al.}(2017){Ponomareva}, {Verheijen}, {Peletier}, \&
  {Bosma}}]{Ponomareva2017}
{Ponomareva}, A.~A., {Verheijen}, M.~A.~W., {Peletier}, R.~F., \& {Bosma}, A.
  2017, \mnras, 469, 2387, \dodoi{10.1093/mnras/stx1018}

\bibitem[{{Prugniel} \& {Simien}(1994)}]{Prugniel&Simien1994}
{Prugniel}, P., \& {Simien}, F. 1994, \aap, 282, L1

\bibitem[{{Prugniel} \& {Simien}(1996)}]{Prugniel&Simien1996}
---. 1996, \aap, 309, 749

\bibitem[{{Randriamampandry} {et~al.}(2015){Randriamampandry}, {Combes},
  {Carignan}, \& {Deg}}]{Randriamampandry2015MNRAS}
{Randriamampandry}, T.~H., {Combes}, F., {Carignan}, C., \& {Deg}, N. 2015,
  \mnras, 454, 3743, \dodoi{10.1093/mnras/stv2147}

\bibitem[{{Renzini} \& {Ciotti}(1993)}]{Renzini1993}
{Renzini}, A., \& {Ciotti}, L. 1993, \apjl, 416, L49, \dodoi{10.1086/187068}

\bibitem[{{Reyes} {et~al.}(2011){Reyes}, {Mandelbaum}, {Gunn}, {Pizagno}, \&
  {Lackner}}]{Reyes2011}
{Reyes}, R., {Mandelbaum}, R., {Gunn}, J.~E., {Pizagno}, J., \& {Lackner},
  C.~N. 2011, \mnras, 417, 2347, \dodoi{10.1111/j.1365-2966.2011.19415.x}

\bibitem[{{Rodr{\'\i}guez} \& {Padilla}(2013)}]{Rodriguez&Padilla2013}
{Rodr{\'\i}guez}, S., \& {Padilla}, N.~D. 2013, \mnras, 434, 2153,
  \dodoi{10.1093/mnras/stt1168}

\bibitem[{{Rodr{\'\i}guez-Puebla} {et~al.}(2015){Rodr{\'\i}guez-Puebla},
  {Avila-Reese}, {Yang}, {Foucaud}, {Drory}, \&
  {Jing}}]{Rodriguez-Puebla2015ApJ}
{Rodr{\'\i}guez-Puebla}, A., {Avila-Reese}, V., {Yang}, X., {et~al.} 2015,
  \apj, 799, 130, \dodoi{10.1088/0004-637X/799/2/130}

\bibitem[{{Roth} {et~al.}(2005){Roth}, {Kelz}, {Fechner}, {Hahn}, {Bauer},
  {Becker}, {B{\"o}hm}, {Christensen}, {Dionies}, {Paschke}, {Popow}, {Wolter},
  {Schmoll}, {Laux}, \& {Altmann}}]{Roth2005PASP}
{Roth}, M.~M., {Kelz}, A., {Fechner}, T., {et~al.} 2005, \pasp, 117, 620,
  \dodoi{10.1086/429877}

\bibitem[{{Salpeter}(1955)}]{Salpeter1955}
{Salpeter}, E.~E. 1955, \apj, 121, 161, \dodoi{10.1086/145971}

\bibitem[{{Sanchez}(2019)}]{Sebastian2019Review}
{Sanchez}, S.~F. 2019, arXiv e-prints, arXiv:1911.06925.
\newblock \doarXiv{1911.06925}

\bibitem[{{S{\'a}nchez} {et~al.}(2012){S{\'a}nchez}, {Kennicutt}, {Gil de Paz},
  {van de Ven}, {V{\'\i}lchez}, {Wisotzki}, {Walcher}, {Mast}, {Aguerri},
  {Albiol-P{\'e}rez}, {Alonso-Herrero}, {Alves}, {Bakos}, {Bart{\'a}kov{\'a}},
  {Bland-Hawthorn}, {Boselli}, {Bomans}, {Castillo-Morales}, {Cortijo-Ferrero},
  {de Lorenzo-C{\'a}ceres}, {Del Olmo}, {Dettmar}, {D{\'\i}az}, {Ellis},
  {Falc{\'o}n-Barroso}, {Flores}, {Gallazzi}, {Garc{\'\i}a-Lorenzo},
  {Gonz{\'a}lez Delgado}, {Gruel}, {Haines}, {Hao}, {Husemann},
  {Igl{\'e}sias-P{\'a}ramo}, {Jahnke}, {Johnson}, {Jungwiert}, {Kalinova},
  {Kehrig}, {Kupko}, {L{\'o}pez-S{\'a}nchez}, {Lyubenova}, {Marino},
  {M{\'a}rmol-Queralt{\'o}}, {M{\'a}rquez}, {Masegosa}, {Meidt},
  {Mendez-Abreu}, {Monreal-Ibero}, {Montijo}, {Mour{\~a}o}, {Palacios-Navarro},
  {Papaderos}, {Pasquali}, {Peletier}, {P{\'e}rez}, {P{\'e}rez}, {Quirrenbach},
  {Rela{\~n}o}, {Rosales-Ortega}, {Roth}, {Ruiz-Lara},
  {S{\'a}nchez-Bl{\'a}zquez}, {Sengupta}, {Singh}, {Stanishev}, {Trager},
  {Vazdekis}, {Viironen}, {Wild}, {Zibetti}, \&
  {Ziegler}}]{SanchezCALIFA2012A&A}
{S{\'a}nchez}, S.~F., {Kennicutt}, R.~C., {Gil de Paz}, A., {et~al.} 2012,
  \aap, 538, A8, \dodoi{10.1051/0004-6361/201117353}

\bibitem[{{S{\'a}nchez} {et~al.}(2016{\natexlab{a}}){S{\'a}nchez},
  {Garc{\'\i}a-Benito}, {Zibetti}, {Walcher}, {Husemann}, {Mendoza}, {Galbany},
  {Falc{\'o}n-Barroso}, {Mast}, {Aceituno}, {Aguerri}, {Alves}, {Amorim},
  {Ascasibar}, {Barrado-Navascues}, {Barrera-Ballesteros}, {Bekerait{\`e}},
  {Bland -Hawthorn}, {Cano D{\'\i}az}, {Cid Fernandes}, {Cavichia}, {Cortijo},
  {Dannerbauer}, {Demleitner}, {D{\'\i}az}, {Dettmar}, {de
  Lorenzo-C{\'a}ceres}, {del Olmo}, {Galazzi}, {Garc{\'\i}a-Lorenzo}, {Gil de
  Paz}, {Gonz{\'a}lez Delgado}, {Holmes}, {Igl{\'e}sias-P{\'a}ramo}, {Kehrig},
  {Kelz}, {Kennicutt}, {Kleemann}, {Lacerda}, {L{\'o}pez Fern{\'a}ndez},
  {L{\'o}pez S{\'a}nchez}, {Lyubenova}, {Marino}, {M{\'a}rquez},
  {Mendez-Abreu}, {Moll{\'a}}, {Monreal-Ibero}, {Ortega Minakata},
  {Torres-Papaqui}, {P{\'e}rez}, {Rosales-Ortega}, {Roth},
  {S{\'a}nchez-Bl{\'a}zquez}, {Schilling}, {Spekkens}, {Vale Asari}, {van den
  Bosch}, {van de Ven}, {Vilchez}, {Wild}, {Wisotzki}, {Y{\i}ld{\i}r{\i}m}, \&
  {Ziegler}}]{Sanches3DR2016A&A}
{S{\'a}nchez}, S.~F., {Garc{\'\i}a-Benito}, R., {Zibetti}, S., {et~al.}
  2016{\natexlab{a}}, \aap, 594, A36, \dodoi{10.1051/0004-6361/201628661}

\bibitem[{{S{\'a}nchez} {et~al.}(2016{\natexlab{b}}){S{\'a}nchez}, {P{\'e}rez},
  {S{\'a}nchez-Bl{\'a}zquez}, {Gonz{\'a}lez}, {Ros{\'a}lez-Ortega},
  {Cano-D{\'{\i}} az}, {L{\'o}pez-Cob{\'a}}, {Marino}, {Gil de Paz},
  {Moll{\'a}}, {L{\'o}pez-S{\'a}nchez}, {Ascasibar}, \&
  {Barrera-Ballesteros}}]{Sanchez2016a}
{S{\'a}nchez}, S.~F., {P{\'e}rez}, E., {S{\'a}nchez-Bl{\'a}zquez}, P., {et~al.}
  2016{\natexlab{b}}, \rmxaa, 52, 21.
\newblock \doarXiv{1509.08552}

\bibitem[{{S{\'a}nchez} {et~al.}(2016{\natexlab{c}}){S{\'a}nchez}, {P{\'e}rez},
  {S{\'a}nchez-Bl{\'a}zquez}, {Garc{\'{\i}}a-Benito}, {Ibarra-Mede},
  {Gonz{\'a}lez}, {Rosales-Ortega}, {S{\'a}nchez-Menguiano}, {Ascasibar},
  {Bitsakis}, {Law}, {Cano-D{\'{\i}}az}, {L{\'o}pez-Cob{\'a}}, {Marino}, {Gil
  de Paz}, {L{\'o}pez-S{\'a}nchez}, {Barrera-Ballesteros}, {Galbany}, {Mast},
  {Abril-Melgarejo}, \& {Roman-Lopes}}]{Sanchez2016b}
---. 2016{\natexlab{c}}, \rmxaa, 52, 171.
\newblock \doarXiv{1602.01830}

\bibitem[{{S{\'a}nchez} {et~al.}(2018){S{\'a}nchez}, {Avila-Reese},
  {Hernandez-Toledo}, {Cortes-Su{\'a}rez}, {Rodr{\'{\i}}guez-Puebla},
  {Ibarra-Medel}, {Cano-D{\'{\i}}az}, {Barrera-Ballesteros}, {Negrete},
  {Calette}, {de Lorenzo-C{\'a}ceres}, {Ortega-Minakata}, {Aquino},
  {Valenzuela}, {Clemente}, {Storchi-Bergmann}, {Riffel}, {Schimoia}, {Riffel},
  {Rembold}, {Brownstein}, {Pan}, {Yates}, {Mallmann}, \&
  {Bitsakis}}]{Sanchez2018AGN}
{S{\'a}nchez}, S.~F., {Avila-Reese}, V., {Hernandez-Toledo}, H., {et~al.} 2018,
  \rmxaa, 54, 217.
\newblock \doarXiv{1709.05438}

\bibitem[{{Schaye} {et~al.}(2015){Schaye}, {Crain}, {Bower}, {Furlong},
  {Schaller}, {Theuns}, {Dalla Vecchia}, {Frenk}, {McCarthy}, {Helly},
  {Jenkins}, {Rosas-Guevara}, {White}, {Baes}, {Booth}, {Camps}, {Navarro},
  {Qu}, {Rahmati}, {Sawala}, {Thomas}, \& {Trayford}}]{Schaye2015MNRAS}
{Schaye}, J., {Crain}, R.~A., {Bower}, R.~G., {et~al.} 2015, \mnras, 446, 521,
  \dodoi{10.1093/mnras/stu2058}

\bibitem[{{Schwarzschild}(1979)}]{Schwarzschild1979}
{Schwarzschild}, M. 1979, \apj, 232, 236, \dodoi{10.1086/157282}

\bibitem[{{Sellwood} \& {S{\'a}nchez}(2010)}]{Sellwood2010}
{Sellwood}, J.~A., \& {S{\'a}nchez}, R.~Z. 2010, \mnras, 404, 1733,
  \dodoi{10.1111/j.1365-2966.2010.16430.x}

\bibitem[{{Smee} {et~al.}(2013){Smee}, {Gunn}, {Uomoto}, {Roe}, {Schlegel},
  {Rockosi}, {Carr}, {Leger}, {Dawson}, {Olmstead}, {Brinkmann}, {Owen},
  {Barkhouser}, {Honscheid}, {Harding}, {Long}, {Lupton}, {Loomis}, {Anderson},
  {Annis}, {Bernardi}, {Bhardwaj}, {Bizyaev}, {Bolton}, {Brewington}, {Briggs},
  {Burles}, {Burns}, {Castander}, {Connolly}, {Davenport}, {Ebelke}, {Epps},
  {Feldman}, {Friedman}, {Frieman}, {Heckman}, {Hull}, {Knapp}, {Lawrence},
  {Loveday}, {Mannery}, {Malanushenko}, {Malanushenko}, {Merrelli}, {Muna},
  {Newman}, {Nichol}, {Oravetz}, {Pan}, {Pope}, {Ricketts}, {Shelden},
  {Sandford}, {Siegmund}, {Simmons}, {Smith}, {Snedden}, {Schneider},
  {SubbaRao}, {Tremonti}, {Waddell}, \& {York}}]{Smee2013}
{Smee}, S.~A., {Gunn}, J.~E., {Uomoto}, A., {et~al.} 2013, \aj, 146, 32,
  \dodoi{10.1088/0004-6256/146/2/32}

\bibitem[{{Spekkens} \& {Sellwood}(2007)}]{Spekkens2007}
{Spekkens}, K., \& {Sellwood}, J.~A. 2007, \apj, 664, 204,
  \dodoi{10.1086/518471}

\bibitem[{{Tonini} {et~al.}(2014){Tonini}, {Jones}, {Mould}, {Webster},
  {Danilovich}, \& {Ozbilgen}}]{Tonini2014}
{Tonini}, C., {Jones}, D.~H., {Mould}, J., {et~al.} 2014, \mnras, 438, 3332,
  \dodoi{10.1093/mnras/stt2442}

\bibitem[{{Trayford} \& {Schaye}(2019)}]{Trayford2019}
{Trayford}, J.~W., \& {Schaye}, J. 2019, \mnras, 485, 5715,
  \dodoi{10.1093/mnras/stz757}

\bibitem[{{Trujillo} {et~al.}(2004){Trujillo}, {Burkert}, \&
  {Bell}}]{Trujillo2004}
{Trujillo}, I., {Burkert}, A., \& {Bell}, E.~F. 2004, \apjl, 600, L39,
  \dodoi{10.1086/381528}

\bibitem[{{Tully} \& {Fisher}(1977)}]{Tully-Fisher1977}
{Tully}, R.~B., \& {Fisher}, J.~R. 1977, \aap, 54, 661

\bibitem[{{{\"U}bler} {et~al.}(2017){{\"U}bler}, {F{\"o}rster Schreiber},
  {Genzel}, {Wisnioski}, {Wuyts}, {Lang}, {Naab}, {Burkert}, {van Dokkum},
  {Tacconi}, {Wilman}, {Fossati}, {Mendel}, {Beifiori}, {Belli}, {Bender},
  {Brammer}, {Chan}, {Davies}, {Fabricius}, {Galametz}, {Lutz}, {Momcheva},
  {Nelson}, {Saglia}, {Seitz}, \& {Tadaki}}]{Ubler2017ApJ}
{{\"U}bler}, H., {F{\"o}rster Schreiber}, N.~M., {Genzel}, R., {et~al.} 2017,
  \apj, 842, 121, \dodoi{10.3847/1538-4357/aa7558}

\bibitem[{{Valenzuela} {et~al.}(2007){Valenzuela}, {Rhee}, {Klypin},
  {Governato}, {Stinson}, {Quinn}, \& {Wadsley}}]{Valenzuela2007}
{Valenzuela}, O., {Rhee}, G., {Klypin}, A., {et~al.} 2007, \apj, 657, 773,
  \dodoi{10.1086/508674}

\bibitem[{{van den Bosch} {et~al.}(2008){van den Bosch}, {van de Ven},
  {Verolme}, {Cappellari}, \& {de Zeeuw}}]{Remco2008MNRAS}
{van den Bosch}, R.~C.~E., {van de Ven}, G., {Verolme}, E.~K., {Cappellari},
  M., \& {de Zeeuw}, P.~T. 2008, \mnras, 385, 647,
  \dodoi{10.1111/j.1365-2966.2008.12874.x}

\bibitem[{{Verheijen}(2001)}]{Verheijen2001}
{Verheijen}, M.~A.~W. 2001, \apj, 563, 694, \dodoi{10.1086/323887}

\bibitem[{{Vogt} {et~al.}(2004){Vogt}, {Haynes}, {Herter}, \&
  {Giovanelli}}]{Vogt2004}
{Vogt}, N.~P., {Haynes}, M.~P., {Herter}, T., \& {Giovanelli}, R. 2004, \aj,
  127, 3273, \dodoi{10.1086/420701}

\bibitem[{{Wake} {et~al.}(2017){Wake}, {Bundy}, {Diamond-Stanic}, {Yan},
  {Blanton}, {Bershady}, {S{\'a}nchez-Gallego}, {Drory}, {Jones}, {Kauffmann},
  {Law}, {Li}, {MacDonald}, {Masters}, {Thomas}, {Tinker}, {Weijmans}, \&
  {Brownstein}}]{Wake2017}
{Wake}, D.~A., {Bundy}, K., {Diamond-Stanic}, A.~M., {et~al.} 2017, \aj, 154,
  86, \dodoi{10.3847/1538-3881/aa7ecc}

\bibitem[{{Walcher} {et~al.}(2014){Walcher}, {Wisotzki}, {Bekerait{\'e}},
  {Husemann}, {Iglesias-P{\'a}ramo}, {Backsmann}, {Barrera Ballesteros},
  {Catal{\'a}n-Torrecilla}, {Cortijo}, {del Olmo}, {Garcia Lorenzo},
  {Falc{\'o}n-Barroso}, {Jilkova}, {Kalinova}, {Mast}, {Marino},
  {M{\'e}ndez-Abreu}, {Pasquali}, {S{\'a}nchez}, {Trager}, {Zibetti},
  {Aguerri}, {Alves}, {Bland-Hawthorn}, {Boselli}, {Castillo Morales}, {Cid
  Fernandes}, {Flores}, {Galbany}, {Gallazzi}, {Garc{\'{\i}}a-Benito}, {Gil de
  Paz}, {Gonz{\'a}lez-Delgado}, {Jahnke}, {Jungwiert}, {Kehrig}, {Lyubenova},
  {M{\'a}rquez Perez}, {Masegosa}, {Monreal Ibero}, {P{\'e}rez}, {Quirrenbach},
  {Rosales-Ortega}, {Roth}, {Sanchez-Blazquez}, {Spekkens}, {Tundo}, {van de
  Ven}, {Verheijen}, {Vilchez}, \& {Ziegler}}]{Walcher2014A&A}
{Walcher}, C.~J., {Wisotzki}, L., {Bekerait{\'e}}, S., {et~al.} 2014, \aap,
  569, A1, \dodoi{10.1051/0004-6361/201424198}

\bibitem[{{Weiner} {et~al.}(2006){Weiner}, {Willmer}, {Faber}, {Melbourne},
  {Kassin}, {Phillips}, {Harker}, {Metevier}, {Vogt}, \& {Koo}}]{Weiner2006}
{Weiner}, B.~J., {Willmer}, C.~N.~A., {Faber}, S.~M., {et~al.} 2006, \apj, 653,
  1027, \dodoi{10.1086/508921}

\bibitem[{{Zaritsky} {et~al.}(2006){Zaritsky}, {Gonzalez}, \&
  {Zabludoff}}]{Zaritsky2006}
{Zaritsky}, D., {Gonzalez}, A.~H., \& {Zabludoff}, A.~I. 2006, \apj, 638, 725,
  \dodoi{10.1086/498672}

\bibitem[{{Zaritsky} {et~al.}(2008){Zaritsky}, {Zabludoff}, \&
  {Gonzalez}}]{Zaritsky2008}
{Zaritsky}, D., {Zabludoff}, A.~I., \& {Gonzalez}, A.~H. 2008, \apj, 682, 68,
  \dodoi{10.1086/529577}

\bibitem[{{Zaritsky} {et~al.}(2011){Zaritsky}, {Zabludoff}, \&
  {Gonzalez}}]{Zaritsky2012}
---. 2011, \apj, 727, 116, \dodoi{10.1088/0004-637X/727/2/116}

\bibitem[{{Zaritsky} {et~al.}(2012){Zaritsky}, {Zabludoff}, \&
  {Gonzalez}}]{Zaritsky2012ApJDistances}
---. 2012, \apj, 748, 15, \dodoi{10.1088/0004-637X/748/1/15}

\bibitem[{{Zaritsky} {et~al.}(2014){Zaritsky}, {Courtois}, {Mu{\~n}oz-Mateos},
  {Sorce}, {Erroz-Ferrer}, {Comer{\'o}n}, {Gadotti}, {Gil de Paz}, {Hinz},
  {Laurikainen}, {Kim}, {Laine}, {Men{\'e}ndez-Delmestre}, {Mizusawa}, {Regan},
  {Salo}, {Seibert}, {Sheth}, {Athanassoula}, {Bosma}, {Cisternas}, {Ho}, \&
  {Holwerda}}]{Zaritsky2014}
{Zaritsky}, D., {Courtois}, H., {Mu{\~n}oz-Mateos}, J.-C., {et~al.} 2014, \aj,
  147, 134, \dodoi{10.1088/0004-6256/147/6/134}

\bibitem[{{Zhu} {et~al.}(2018{\natexlab{a}}){Zhu}, {van de Ven},
  {M{\'e}ndez-Abreu}, \& {Obreja}}]{Zhuq062018}
{Zhu}, L., {van de Ven}, G., {M{\'e}ndez-Abreu}, J., \& {Obreja}, A.
  2018{\natexlab{a}}, \mnras, 479, 945, \dodoi{10.1093/mnras/sty1521}

\bibitem[{{Zhu} {et~al.}(2018{\natexlab{b}}){Zhu}, {van den Bosch}, {van de
  Ven}, {Lyubenova}, {Falc{\'o}n-Barroso}, {Meidt}, {Martig}, {Shen}, {Li},
  {Yildirim}, {Walcher}, \& {Sanchez}}]{Zhu2018}
{Zhu}, L., {van den Bosch}, R., {van de Ven}, G., {et~al.} 2018{\natexlab{b}},
  \mnras, 473, 3000, \dodoi{10.1093/mnras/stx2409}

\bibitem[{{Zhu} {et~al.}(2018{\natexlab{c}}){Zhu}, {van de Ven}, {van den
  Bosch}, {Rix}, {Lyubenova}, {Falc{\'o}n-Barroso}, {Martig}, {Mao}, {Xu},
  {Jin}, {Obreja}, {Grand }, {Dutton}, {Macci{\`o}}, {G{\'o}mez}, {Walcher},
  {Garc{\'\i}a-Benito}, {Zibetti}, \& {S{\'a}nchez}}]{Zhu2018b}
{Zhu}, L., {van de Ven}, G., {van den Bosch}, R., {et~al.} 2018{\natexlab{c}},
  Nature Astronomy, 2, 233, \dodoi{10.1038/s41550-017-0348-1}

\bibitem[{{Zwaan} {et~al.}(1995){Zwaan}, {van der Hulst}, {de Blok}, \&
  {McGaugh}}]{Zwaan1995}
{Zwaan}, M.~A., {van der Hulst}, J.~M., {de Blok}, W.~J.~G., \& {McGaugh},
  S.~S. 1995, \mnras, 273, L35, \dodoi{10.1093/mnras/273.1.L35}

\bibitem[{{Zwicky}(1933)}]{Zwicky1933}
{Zwicky}, F. 1933, Helvetica Physica Acta, 6, 110

\end{thebibliography}

\end{document}